\RequirePackage{fix-cm}
\documentclass[smallcondensed]{svjour3}     
\smartqed  
\usepackage[numbers]{natbib}
\usepackage{graphicx}
\usepackage{lineno,hyperref}
\modulolinenumbers[5]
\usepackage{amsmath}
\usepackage{balance}
\usepackage[byname]{smartref}
\usepackage{txfonts}
\usepackage{tocloft}
\usepackage{amsmath,amssymb,amsfonts} 
\usepackage{color}                    
\usepackage{comment}
\usepackage{mathtools}
\usepackage{commath}
\usepackage{graphics}
\usepackage[font={small,it}]{caption}
\usepackage{placeins}
\usepackage{tikz}
\usepackage{bchart}
\usepackage{enumitem}
\usepackage{adjustbox}
\usepackage{pstricks,pst-node,pst-tree}
\usepackage{pst-plot}
\usepackage{multirow}
\usepackage{epstopdf}
\AppendGraphicsExtensions{.tif}
\journalname{Applied Intelligence}
\begin{document}

\title{Multi-split Optimized Bagging Ensemble Model Selection for Multi-class Educational Data Mining
}
\subtitle{}


\author{MohammadNoor Injadat        \and
        Abdallah Moubayed 			\and 
        Ali Bou Nassif				\and
        Abdallah Shami 
}


\institute{MohammadNoor Injadat, Abdallah Moubayed, Abdallah Shami \at
              Electrical \& Computer Engineering Dept.\\ 
              University of Western Ontario\\ 
              London, ON, Canada \\
              \email{minjadat@uwo.ca, amoubaye@uwo.ca, abdallah.shami@uwo.ca}           
            \and
            Ali Bou Nassif \at
            Computer Engineering Dept.\\ 
            University of Sharjah, Sharjah, UAE\\ 
            and\\
            Electrical \& Computer Engineering Dept.\\ 
            University of Western Ontario\\ 
            London, ON, Canada \\
            \email{anassif@sharjah.ac.ae}
}

\date{Received: date / Accepted: date}

\maketitle

\begin{abstract}
Predicting students' academic performance has been a research area of interest in recent years with many institutions focusing on improving the students' performance and the education quality. The analysis and prediction of students' performance can be achieved using various data mining techniques. Moreover, such techniques allow instructors to determine possible factors that may affect the students' final marks. To that end, this work analyzes two different undergraduate datasets at two different universities. Furthermore, this work aims to predict the students' performance at two stages of course delivery (20\% and 50\% respectively). This analysis allows for properly choosing the appropriate machine learning algorithms to use as well as optimize the algorithms' parameters. Furthermore, this work adopts a systematic multi-split approach based on Gini index and p-value. This is done by optimizing a suitable bagging ensemble learner that is built from any combination of six potential base machine learning algorithms. It is shown through experimental results that the posited bagging ensemble models achieve high accuracy for the target group for both datasets. 
\end{abstract}

\section{Introduction}\label{sec:multi_intro}
\indent Data mining is rapidly becoming a part of software engineering projects, and standard methods are constantly revisited to integrate the software engineering point of view. Data mining can be defined as an extraction of data from a dataset and discovering useful information from it \cite{1}\cite{s12}. This is followed by the analysis of the collected data in order to enhance the decision-making process \cite{2}. Data mining uses different algorithms and tries to uncover certain patterns from data \cite{3}. These techniques techniques have proved to be effective solutions in a variety of fields including education, network security, and business \cite{ff,se,sd}. Hence, they have the potential to also be effective in other fields such as medicine and education. \\
Educational Data Mining (EDM), a sub-field of data mining, has emerged that specializes in educational data with the goal of better understanding students’ behavior and improving their performance \cite{4}\cite{springer1}. Moreover, this sub-field also aims at enhancing the learning and teaching processes \cite{2}. EDM often takes into consideration various types of data such as administrative data, students’ performance data, and student activity data to gain insights and provide the appropriate recommendation \cite{5} \cite{springer2}.\\
The rapid growth of technology and the Internet has introduced an interactive opportunities to help education field to improve the teaching and learning processes. In turn, this has led to the emergence of the field of e-learning. This field can be defined as “the use of computer network technology, primarily over an intranet or through the Internet, to deliver information and instruction to individuals” \cite{7}\cite{springer3}. There are various challenges facing e-learning platforms and environment \cite{s11}. This includes the assorted styles of learning, and challenges arising from cultural differences \cite{9}. Other challenges also exist such as pedagogical e-learning, technological and technical training, and e-learning time management \cite{10}. To this end, personalized learning has emerged as a necessity in order to better cater to the learners' needs \cite{injadat_ch4}. Accordingly, this personalization process has become a challenging task \cite{11}, as it requires adapting courses to meet different individuals' needs. This calls for adaptive techniques to be implemented \cite{12}, \cite{13}. This can be done by automatically collecting data from the e-learning environment \cite{13} and analyzing the learner’s profile to customize the course according to the participant’s needs and constraints such as his/her location, language, currency, seasons, etc. \cite{13}, \cite{14}, \cite{15}.\\
Many of the previous works in the literature focused on predicting the performance of the students by adopting a binary classification model. However, some educators prefer to identify not only two classes of students (i.e. Good vs. Weak), but instead they divide the students into several groups and consider the associated multi-class classification problem \cite{ch5_D0}. This is usually done because the binary model often identifies a large number weak students, many of which are not truly at risk of failing the course. Accordingly, this work considers two datasets at two different stages of the course, namely at 20\% and 50\% of the coursework, and divides the students into three groups, namely Weak, Fair, and Good students. Accordingly, the datasets are analyzed as a set of multi-class classification problems.\\
Multi-class classification problems can be solved by naturally extending the binary classification techniques for some algorithms, \cite{ch5_D}. In this work, we consider various classification algorithms, compare their performances, and use Machine Learning (ML) techniques aiming to predict the students' performance in the most accurate way. Indeed, we consider K-nearest neighbor (k-NN), random forest (RF), Support Vector Machine (SVM), Multinomial Logistic Regression (LR), Naïve Bayes (NB) and Neural Networks (NN) and use an optimized systematic ensemble model selection approach coupled with ML hyper-parameter tuning using grid search optimization.\\  
In this paper, we produced a bagging of each type of model and the bagging was used for the ensembles as opposed to single models. Bagging is itself an ensemble algorithm as it consists of grouping several models of the same type and defining a linear combination of the individual predictions as the final prediction on an external test sample, as explained in Section \ref{sec.method}. Bagging is one of the best procedures to improve the performance of classifiers as it helps reduce the variance in many hard decision problems \cite{ch5_A}\cite{se2}. The empirical fact that bagging improves the classifiers' performance is widely documented \cite{ch5_A2}, and in fact ensemble methods placed first in many prestigious ML competitions, such as the Netflix Competition \cite{ch5_B2}, KDD 2009 \cite{ch5_B3}, and Kaggle \cite{ch5_B}. Furthermore, a multi-split framework is considered for the studied datasets in order to reduce the bias of the ML models investigated as part of the bagging ensemble models. \\  
The main disadvantage of bagging, and other ensemble algorithms, is the lack of interpretation. For instance, a linear combination of decision trees is much harder to interpret than a single tree. In the same way, bagging several variable selections gives little clues about which of the predictor variables are actually important.In this paper, in order to have a rough idea of which variables are the best predictors for each algorithm, we decided to average, for each variable, its importance in every model and this average is assigned to the variable and defined to be its \emph{averaged importance}. This was done in order to better highlight the features that are truly important across the multiple splits under consideration. \\
The remainder of this paper is organized as follows: Section \ref{sec:related_work_ch5} presents some of the previous related work and their limitations; Section \ref{ch5_contribution} summarizes the research contributions of this work; Section \ref{Sec:multi_target} describes the datasets under consideration and defines the corresponding target variables for both datasets; Section \ref{Sec:PerfMeas} describes the performance measurement approach adopted; 
Section \ref{sec.method} presents the methodology used to choose the best classifiers for the multi-class classification problem; Section \ref{ML_multi} discusses the architecture used for training NN and shows the features' importance for each classifier for each dataset; Section \ref{sec:multi_results} presents and discusses the experimental results both in terms of Gini Indices (also called Gini coefficient) and by using confusion matrices; and finally, Section \ref{sec:future2} lists the research limitations, proposes multiple future research opportunities, and concludes the paper.
\section{Related Work and Limitations}\label{sec:related_work_ch5}
\subsection{Related Work}
Educational data mining has become a rich field of research with the demand for empirical studies and research by academia increasing in recent years. This is due to the competitive advantages that can be gained from such kind of research. Data mining can be used to evaluate and analyze the different factors that improve the knowledge gaining, skills improvement of the learners, and makes the educational institution offer a better learning experience with highly qualified students or trainees \cite{ch5re1}.\\
Several researchers have explored the use of data mining techniques in an educational setting. Authors of \cite{ch5re2} used data mining techniques to analyze the learner’s web usage and content-based profiles to have an on-line automatic recommendation system. In contrast, Chang et al. proposed a k-NN classification model to classify the learner’s style \cite{ch5re3}. The results of this model was used to help the educational institution management and faculties to improve the courses’ contents to satisfy the learner’s needs \cite{ch5re3}.\\ 
Another related study that used simple leaner regression to check the effect of the student mother’s education level and the family’s income in learner’s academic level was presented in  \cite{ch5re4}.\\
On the other hand, Baradwaj and Pal used classification methods to evaluate the students’ performance using decision trees \cite{ch5re5}. The study was conducted using collected data from previous year’s database to predict the student result at the end of the current semester. Their study aimed to provide a prediction that will help the next term instructors identify students that they may need help.\\
Other researchers \cite{ch5re6} applied Naïve Bayes classification algorithm to predict students’ grades based on their previous performance and other important factors. The authors discovered that, other than students’ efforts, factors such as residency, the qualification standards of the mother, hobbies and activities, the total income of the family, and the state of the family had a significant effect on the students’ performance. \\
Later, the same authors used Iterative Dichotomiser 3 (ID3) decision tree algorithm and if-then rules to accurately predict the performance of the students at the end of the semester \cite{ch5re7} based on different variables like Previous Semester Marks, Class Test Grades, Seminar Performance, Assignments, Attendance, Lab Work, General Proficiency, and End Semester Marks.\\
Similarly, Moubayed et al. \cite{ch5re8,ch5ref8a} studied the student engagement level using K-means algorithm and derived a set of rulers that related student engagement with academic performance using Apriori association rules algorithm. The results analysis showed a positive correlation between students' engagement level and their academic performance in an e-learning environment.\\
Prasad et al. \cite{ch5re9} used J48 (C4.5) algorithm and concluded that this algorithm is the best choice for making the best decision about the students’ performance. The algorithm was also preferred because of its accuracy and speed. \\
Ahmed and Elaraby conducted a similar research in 2014 \cite{ch5re10} using classification rules. They analyzed data from a course program across 6 years and were able to predict students’ final grades. In similar fashion, Khan et al. \cite{ch5re11} used J48 (C4.5) algorithm for predicting the final grade of Secondary School Students based on their previous marks.\\
Kostiantis et al. \cite{ch5ref11a} proposed an incremental majority voting-based ensemble classifier based on 3 base classifiers, namely NB, k-NN, and Winnow algorithms. The authors' experimental results showed that the proposed ensemble model outperformed the single base models in a binary classification environment.\\
Saxena \cite{ch5re12} used k-means clustering and J48 (C4.5) algorithms and compared their performance in predicting students’ grades. The author concluded that J48 (C4.5) algorithm is more efficient, since it gave higher accuracy values than k-means algorithm. Authors in \cite{ch5re13} used and compared K-Means and Hierarchical clustering algorithms. They concluded that K-means algorithm is more preferred to hierarchical clustering due to better performance and faster model building time.\\
Wang et al. proposed an e-Learning recommendation framework using deep learning neural networks model \cite{ch5re14}. Their experiments showed that the proposed framework offered a better personalized e-learning experience. Similarly, Fok et al. proposed a deep learning model using TensorFlow to predict the performance of students using both academic and non-academic subjects \cite{ch5re15}. Experimental results showed that the proposed model had a high accuracy in terms of student performance prediction.\\
Asogbon \textit{et al.} proposed a multi-class SVM model to correctly predict students’ performance in order to admit them into appropriate faculty program \cite{multi_class_prediction1}. The performance of the model was examined using an educational dataset collected at the University of Lagos, Nigeria. Experimental results showed that the proposed model adequately predicted the performances of students across all categories \cite{multi_class_prediction1}.\\
In a similar fashion, Athani \textit{et al.} also proposed the use of a multi-class SVM model to predict the performance of high school students and classify them into one of five letter grades A-F \cite{multi_class_prediction2}. The goal was to predict student performance to provide a better illustration of the education level of the schools based on their students' failure rate. The authors used a Portuguese high school dataset consisting mostly of the students' socio-economic descriptors as features. Their experiments showed that the proposed multi-class SVM model achieved high prediction accuracy close to 89\% \cite{multi_class_prediction2}.\\
Jain and Solanki proposed a comparative study between four tree-based models to predict the performance of students based on a three-class output \cite{multi_class_prediction3}. Similar to the work of Athani \textit{et al.}, the authors in this work also considered the Portuguese high school dataset consisting mostly of the students' socio-economic descriptors as features. Experimental results showed that the proposed tree-based model also achieved high prediction accuracy with a low execution time \cite{multi_class_prediction3}. 
\subsection{Limitations of Related Work}\label{related_work_limitation}
The limitations of the related work can be summarized as follows:
\begin{itemize}
	\item Do not analyze the features before applying any ML model. Any classification model is directly applied without studying the nature of the data being considered.
	\item Mostly consider the binary classification case. Such cases often lead to identifying too many students which are not truly in danger of failing the course and hence would not need as much help and attention. Even when multi-class models were considered, the features used were mostly focused on students' socio-economic status rather than their performance in different educational tasks. 
	\item Often use a single classification model or an ensemble model built upon randomly chosen group of base classifiers. Moreover, to the best of our knowledge, only majority voting-based ensemble models are considered.
	\item Often predict the performance of students from one course to the other or from one year to the other. Performance prediction is rarely considered during the course delivery.
	\item Often use the default parameters of the utilized algorithms/techniques without optimization.
\end{itemize}
\section{Research Contribution}\label{ch5_contribution}
To overcome the limitations presented in Section \ref{related_work_limitation}, our research aims to predict the students’ performance during the course delivery as opposed to other previous works that perform the prediction at the end of the course. The multi-class classification problem assumes that their is a proportional relationship between the students' efforts and seriousness in the course and their final course performance and grade. \\
More specifically, our work aims to: 
\begin{itemize}
	\item \textit{Analyze} the collected datasets and visualize the corresponding features by applying different graphical and quantitative techniques (e.g. dataset distribution visualization, target variable distribution, and feature importance).
	\item \textit{Optimize} hyper-parameters of the different ML algorithms under consideration using \textit{grid search} algorithm.
	\item \textit{Propose} a systemic approach to build a multi-split-based (to reduce bias) bagging ensemble (to reduce variance) learner to select the most suitable model depending on multiple performance metrics, namely the Gini index (for better statistical significance and robustness) and the target class score.
	\item \textit{Study} the performance of the proposed ensemble learning classification model on multi-class datasets.
	\item \textit{Evaluate} the performance of the proposed bagging ensemble learner in comparison with classical classification techniques.
\end{itemize}
 Note that in this work, the term \textit{Gini index} refers to the Gini coefficient that is calculated based on the Lorenz curve and area under the curve terms \cite{Z}. Therefore, the remainder of this work adopts to the term \textit{Gini index}.
\section{Dataset and Target Variable Description} \label{Sec:multi_target}
\subsection{Dataset Description}
In this section,the two datasets under consideration are described at the two course delivery stages (20\% and 50\% of the coursework). This corresponds to the results of a series of tasks performed by University students. Moreover, Principal Components Analysis (PCA) is conducted to better visualize the considered datasets. 
\begin{itemize}
	\item \emph{Dataset 1}:
	The experiment was conducted at the University of Genoa on a group of 115 first year engineering major students \cite{67}. The dataset consists of data collected using a simulation environment named Deeds (Digital Electronics Education and Design Suite). This e-Learning platform allows students to access the courses’ contents using a special browser and asks the students to solve problems that are distributed over different complexity levels.\\ 
	Table \ref{tab:table_dataset_1_list} shows a summary of the different tasks for which the data was collected. It is worth mentioning that 52 students out of the original 115 students registered were able to complete the course. \\
	The 20\% stage consists of the grades of tasks ES 1.1 to ES 3.5. On the other hand, the 50\% stage consists of tasks ES. 1.1 to ES 5.1.\\ 
	To improve the accuracy of the classification model, empty marks were replaced with a 0. Moreover, all tasks' marks were converted to a scale out of 100. Furthermore, all decimal point marks were rounded to the nearest 1 to maintain consistency. 
	\begin{table}[h!]
		\centering
		\caption{Dataset 1 - Features}
		\resizebox{0.6\textwidth}{!}{
			\begin{tabular}{|c|c|c|c|} 
				\hline     
				\textbf{Feature} & \textbf{Description} & \textbf{Type} & \textbf{Value/s} \\
				\hline
				\text{Id}           &\text{Student Id.} &\text{Nominal} & \text{Std. 1,..,Std. 52} \\
				\hline
				\text{ES 1.1}    & \text{Exc. 1.1 Mark} &\text{Numeric} & \text{0..2} \\
				\hline
				\text{ES 1.2}    & \text{Exc. 1.2 Mark} &\text{Numeric} & \text{0..3} \\
				\hline
				\text{ES 2.1}    & \text{Exc. 2.1 Mark} &\text{Numeric} & \text{0..2} \\
				\hline
				\text{ES 2.2}     & \text{Exc. 2.2 Mark} &\text{Numeric} & \text{0..3} \\
				\hline
				\text{ES 3.1}     & \text{Exc. 3.1 Mark} &\text{Numeric} & \text{0..1} \\
				\hline
				\text{ES 3.2}     &\text{Exc. 3.2 Mark} &\text{Numeric} & \text{0..2} \\
				\hline
				\text{ES 3.3}        & \text{Exc. 3.3 Mark} &\text{Numeric} & \text{0..2} \\
				\hline
				\text{ES 3.4}        & \text{Exc. 3.4 Mark} &\text{Numeric} & \text{0..2} \\
				\hline
				\text{ES 3.5} &\text{Exc. 3.5 Mark} &\text{Numeric} & \text{0..3} \\
				\hline
				\text{ES 4.1} &\text{Exc. 4.1 Mark} &\text{Numeric} & \text{0..15} \\
				\hline
				\text{ES 4.2} & \text{Exc. 4.2 Mark} &\text{Numeric} & \text{0..10} \\
				\hline
				\text{ES 5.1} & \text{Exc. 5.1 Mark} &\text{Numeric} & \text{0..2} \\
				\hline
				\text{ES 5.2} &\text{Exc. 5.2 Mark} &\text{Numeric} & \text{0..10} \\
				\hline
				\text{ES 5.3} & \text{Exc. 5.3 Mark} &\text{Numeric} & \text{0..3} \\
				\hline
				\text{ES 6.1} & \text{Exc. 6.1 Mark} &\text{Numeric} & \text{0..25} \\
				\hline
				\text{ES 6.2} & \text{Exc. 6.2 Mark} &\text{Numeric} & \text{0..15} \\
				\hline
				\text{Final Grade}&  \text{Total Final Mark} &\text{Numeric} & \text{0..100} \\
				\hline
				\text{Total} &\text{Final Course Grade} &\text{Nominal} & \text{G,F,W} \\
				\hline
		\end{tabular}}
		\label{tab:table_dataset_1_list}
	\end{table}
	\item \emph{Dataset 2}:
	This dataset was collected at the University of Western Ontario for a second year undergraduate Science course. The dataset is composed of two main parts. The first part is an event log of the 486 students enrolled. This event log dataset consists of 305933 records. In contrast, the other part, which is under consideration in this research, is the grades of the 486 students in the different evaluated tasks. This includes assignments, quizzes, and exams. \\
	Table \ref{tab:table_dataset_2} summarizes the different tasks evaluated within this course. The 20\% stage consists of the results of Assignment 01 and Quiz 01. On the other hand, the 50\% stage consists of the grades of Quiz 01, Assignments 01 and 02, and the midterm exam. \\ 
	Similar to Dataset 1, all empty marks were replaced with a value of 0 for better classification accuracy. Moreover, all marks were scaled out of 100. Additionally, decimal point marks were rounded to the nearest 1. 
	
	\begin{table}[t]
		\centering
		\caption{ Dataset 2 - Features}
		\scalebox{0.9}{ 
			\begin{tabular}{|l|p{2.5cm}|c|c|} 
				\hline     
				\textbf{Feature} & \textbf{Description} & \textbf{Type} & \textbf{Value/s} \\
				\hline
				\text{Id}           &\text{Student Id.} &\text{Nominal} & \text{std000,..,std485} \\
				\hline
				\text{Quiz01}    & \text{Quiz1 Mark} &\text{Numeric} & \text{0..10} \\
				\hline
				\text{Assign.01}    & \text{Assign.01 Mark} &\text{Numeric} & \text{0..8} \\
				\hline
				\text{Midterm}    & \text{Midterm Mark} &\text{Numeric} & \text{0..20} \\
				\hline
				\text{Assign.02}     & \text{Assign.02 Mark} &\text{Numeric} & \text{0..12} \\
				\hline
				\text{Assign.03}     & \text{Assign.03 Mark} &\text{Numeric} & \text{0..25} \\
				\hline
				\text{Final Exam} & \text{Final Exam Mark} &\text{Numeric} & \text{0..35} \\
				\hline
				\text{Final Grade}&  \text{Total Final Mark} &\text{Numeric} & \text{0..100} \\
				\hline
				\text{Total} &\text{Final Grade} &\text{Nominal} & \text{G,F,W} \\
				\hline
			\end{tabular}
		}
		\label{tab:table_dataset_2}
	\end{table}
\end{itemize}
\subsection{Target Variable Description}
\begin{figure}[!h]
	\centering
	\includegraphics[scale=0.6]{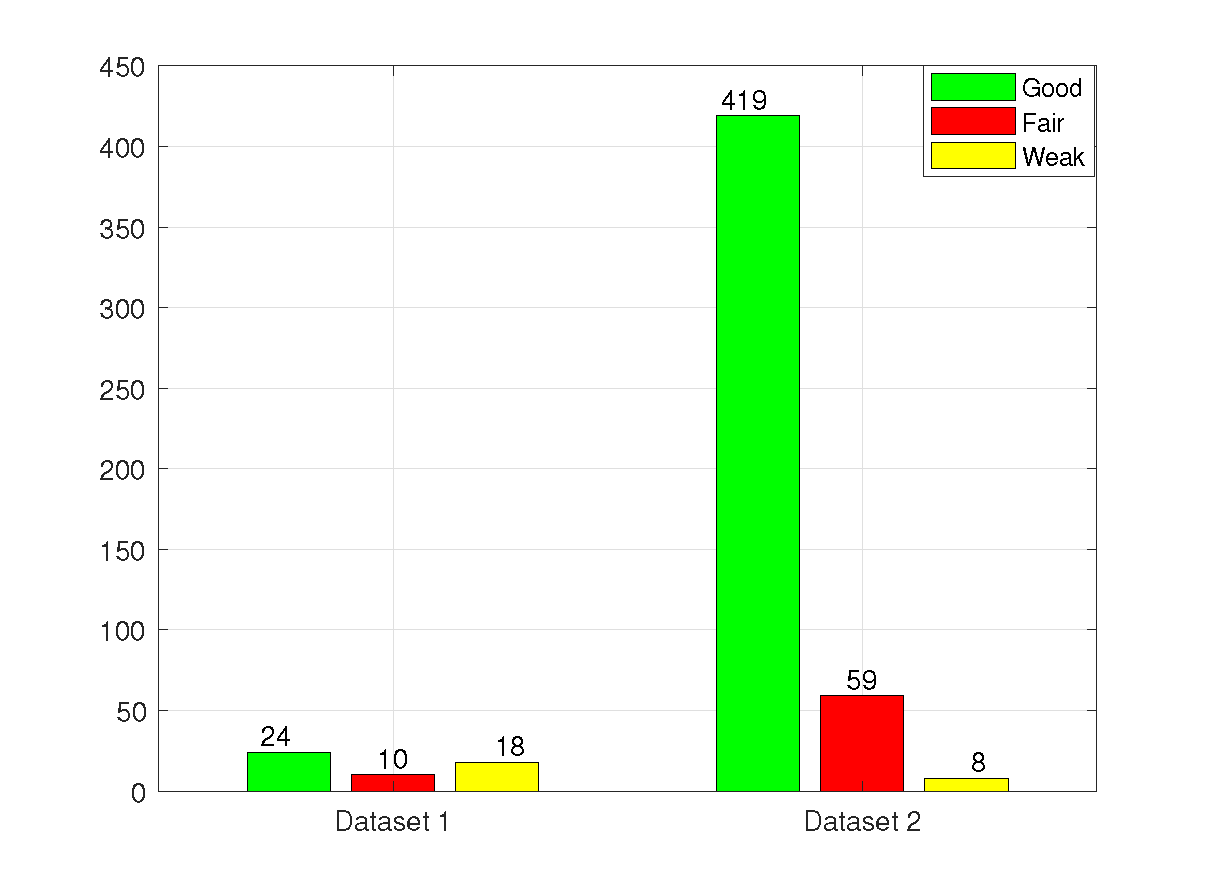}
	\caption{\label{Dataset1_and2Target_Multi}{Dataset 1 and Dataset 2- Target Variables}}
\end{figure}

For the two datasets under consideration, the target variables were constructed by considering the final grade. More specifically, the students were grouped into three groups as follows:
\begin{enumerate} 
	\item Good (G) – the student will finish the course with a good grade ($70-100\%$);
	\item Fair (F) – the student will finish the course with a fair grade ($51-69\%$);
	\item Weak (W) – the student will finish the course with a weak grade ($\leq 50\%$).
\end{enumerate}
In this case, the target group is the Weak students (W) who are predicted to receive a mark below 50\%, meaning that they are at risk of failing the course.
Figure \ref{Dataset1_and2Target_Multi} shows that Datasets 1 and 2 are characterized by being small sized and unbalanced respectively. These two issues have more of an impact on the classification problem. It can be seen that for the first dataset, the three classes are relatively evenly distributed, but each class consists of only a few students. On the other hand, the second dataset is not small sized but is strongly unbalanced, having only 8 Weak students out of 486 students.

\begin{figure}[!h]
	\centering
	\includegraphics[scale=0.5]{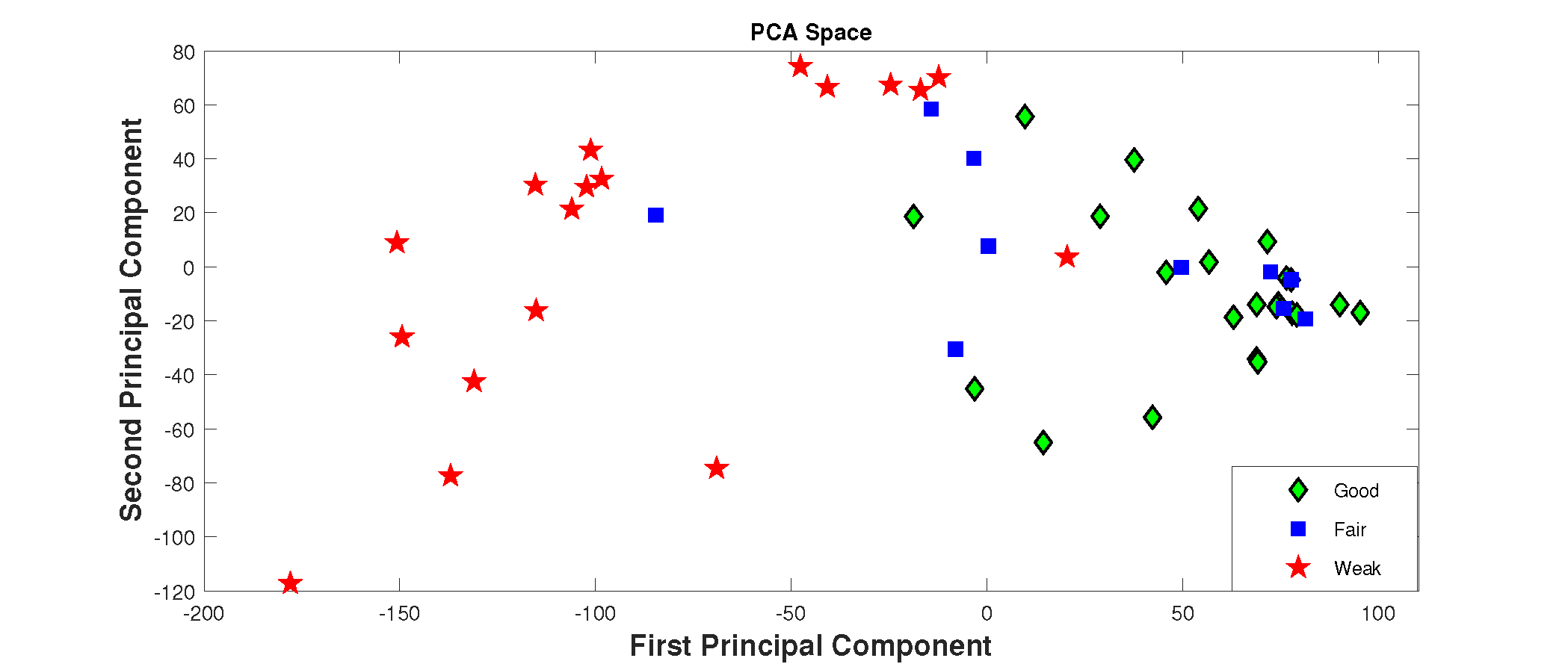}
	\caption{\label{Dataset1_multiPCA}{Dataset 1 - multi-class target visualization}}
\end{figure}

\begin{figure}[!h]
	\centering
	\includegraphics[scale=0.5]{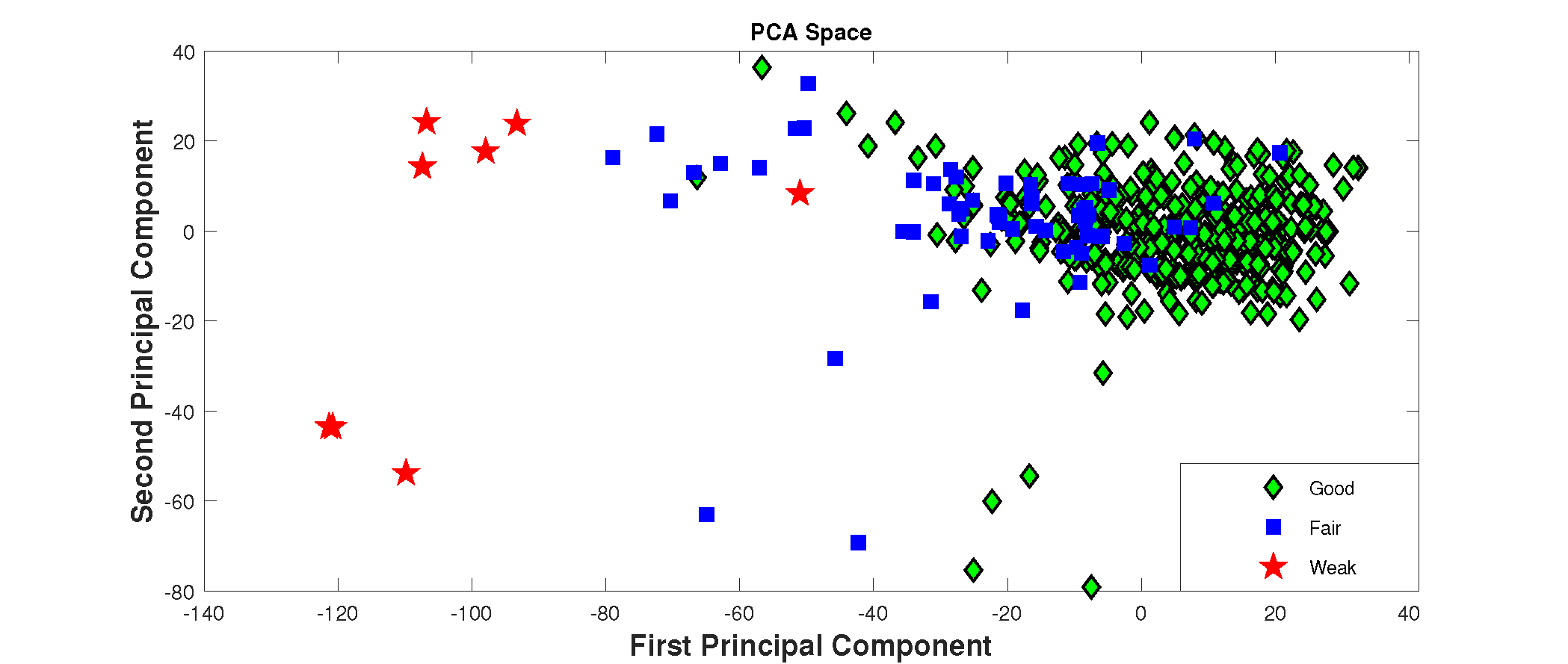}
	\caption{\label{Dataset2_multiPCA}{Dataset 2 - multi-class target visualization}}
\end{figure}

To better visualize the three classes, we applied PCA to the datasets (both considered at Stage 50\%) as shown in Figures \ref{Dataset1_multiPCA} and \ref{Dataset2_multiPCA}. Looking at these two figures, we note that it can be possible to draw a boundary that separates Weak Students from the rest of the students, whereas Fair and Good students are too close and not separable by a boundary. We will see in the next sections that the performance of the models is affected by this distribution and that most of the algorithms fail in distinguishing between Fair and Good students, especially for Dataset 1.

\section{Performance Evaluation Metrics Description}\label{Sec:PerfMeas}
In general there are two standard approaches to choosing multiple class performance measures \cite{ch5_D}, \cite{ch5_C}. One approach, namely \emph{OVA (One-versus-all)}, is to reduce the problem of classifying among $N$ classes into $N$ binary problems. In this case, every class is discriminated from the other classes. In the second approach, called \emph{AVA (All-versus-all)}, each class is compared to each other class. In other words, it is necessary to build a classifier for every pair of classes, i.e. building $\frac{N(N-1)}{2}$ classifiers, while discarding the rest of the classes.\\
Due to the size of our datasets, we chose to follow the first method as opposed to the second one. In fact, if we were to use the second approach for Dataset 1, we would need to train three binary models, one for each pair of classes (G,F), (F,W), and (G,W). In particular, the subset of data for the (F,W) model would consist of only 28 students, which would be split into Training Sample (70\%) and Test Sample (30\%). This corresponds to training a model using 20 students and testing it using only 8 students. Due to the relatively small size of the (F,W) model, we determine that the AVA approach would not be suitabe for accurate prediction.\\
It is well-known that the Gini Index metric, as well as the other metrics (Accuracy, ROC  curve etc.) can be generalized to the multi-class classification problem.
In particular, \emph{we choose the Gini Index metric instead of the Accuracy because the latter depends on the choice of a threshold whereas the Gini Index metric does not}. This makes it statistically more significant and robust than the accuracy, particularly given that it provides a measure of the statistical dispersion of the classes \cite{gini_reason}.\\
In particular, we implemented a generalization of Gini index metric: during the training phase, that computes the Gini Index of each one of the three binary classifications and \emph{optimizes (i.e. maximizes) the average of the 3 performances}, i.e. the performances corresponding to classes G, F, W.  
\section{Methodology}\label{sec.method}
\begin{figure}[h!]
	\centering
	\includegraphics[scale=0.5]{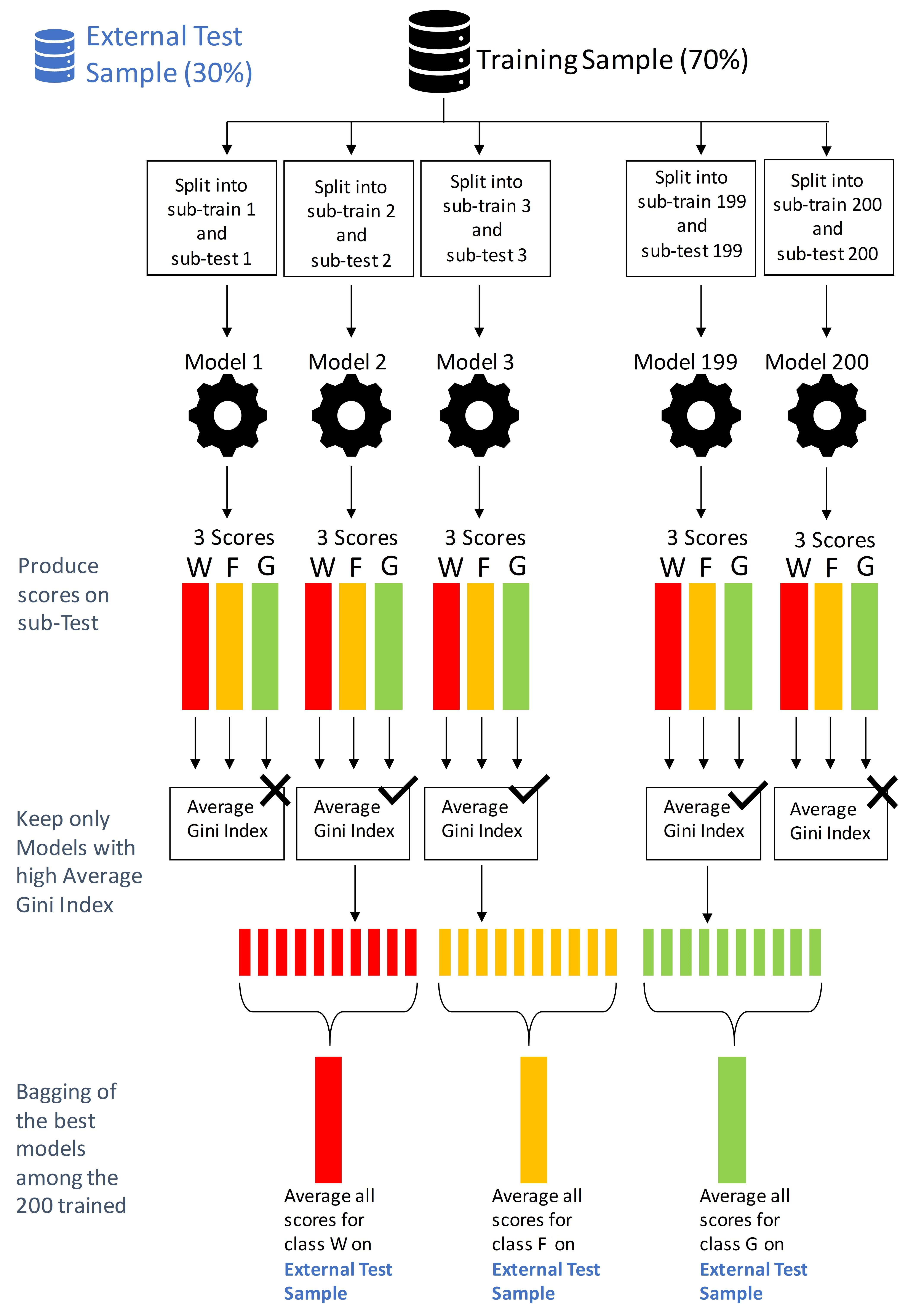}
	\caption{\label{Bagging_explaination}{Bagging Ensemble Model Building Methodology}}
\end{figure}

For the multi-class classification problem we used several algorithms. More specifically we explored RF, SVM - RBF, k-NN, NB, LR, and NN with 1, 2 and 3 layers (i.e. 3 different NN models), for a total of eight classifiers per dataset. \\
In order to achieve better performances, we did not build only one individual model for each algorithm, instead we constructed baggings of classifiers. In fact, as explained in the previous section, bagging reduces the variance.\\
We built a bagging of models for each algorithm in the following way: we started by splitting each dataset into Training and Test samples in proportions 70\%-30\% then we used the training sample to build baggings of models. More precisely the Training sample was split 200 times into sub-Training and sub-Test samples randomly but forcing the percentages of Fair, Good and Weak students to be the same as the ones in the entire datasets.\\
The models resulting from the 200 splits were trained on the sub-Training samples and inferred on the corresponding sub-Test samples. If the Average Gini Index was above a certain fixed threshold (lowest acceptable Gini Index) then the model was kept otherwise it was discarded. For each algorithm we obtained in this way a set of models having the best performances, and we averaged their scores on the (external) Test sample, class by class.  This procedure is explained in Figure \ref{Bagging_explaination}.\\
Once we had the eight baggings of models (one for each algorithm), we considered all the possible ensembles that could be constructed with them and compared their performances in terms of Gini Index, as explained in Section \ref{Sec:PerfMeas}. Moreover, for each dataset, we computed the p-values corresponding to each one of the 256 possible ensembles and aimed to choose as the final ensemble the one that had best Gini Index and, at the same time, that was statistically significant.\\
The Gini Index, also commonly referred to as the Gini coefficient, can be seen geometrically as the area between the Lorenz curve \cite{Z} and the diagonal line representing perfect equality. The higher the Gini Index, the better the performance of the model. Formally the Gini index is defined as follows:\\
Let $F(z)$ be the cumulative distribution of $z$ and let $a$ and $b$ be the highest and the lowest value of $z$ respectively, then the we can calculate half of Gini's expected mean difference as:
\begin{equation} 
2 \int_{a}^{b} F(z)[1-F(z)] dz
\end{equation}
Alternatively, the Gini index can be calculated as 2 $*$ Area Under Curve $-$ 1.\\
On the other hand, the statistical significance of our results is determined by computing the p-values. The general approach is to test the validity of a claim, called the \emph{null hypothesis}, made about a population. An alternative hypothesis is the one you would believe if the null hypothesis is concluded to be untrue. A small p-value ($\leq 0.05$) indicates strong evidence against the null hypothesis,  so you reject the null hypothesis. For our purposes, the null hypothesis states that the Gini Indices were obtained by chance. We generated 1 million random scores from normal distribution  and calculated the p-value. The ensemble learners selected have p-value $\leq 0.05$, indicating that there is strong evidence against the null hypothesis. Therefore, choosing an ensemble model using a combination of Gini Index and p-value allows us to have a more statistically significant and robust model.\\ 
The classifiers were inferred on the test sample, giving as output three vectors of predictions to be analyzed. These three vectors express the chance that each student is classified as Weak, Fair and Good. In order t o build the confusion matrices, we fixed a threshold for each class, namely $\tau_F$, $\tau_G$, and $\tau_W$.  To determine each threshold, a one-vs-all method is considered for each class with the threshold being chosen as the score for which the point on the ROC curve is closest to the top-left corner (commonly referred to as the Youden Index) \cite{optimal_threshold}. This is done in order to find the point that simultaneously maximizes the sensitivity and specificity.\\
For each student belonging to the Test sample, we defined the predicted class according to the following steps: 
\begin{enumerate}
	\item The 3 scores corresponding to the 3 classes were normalized in order to make them comparable.
	\item For each class, if the probability is higher than the corresponding threshold then the target variable for the binary classification problem associated to that class is predicted to be 1, otherwise it's 0. 
	\item In this way we obtained a 3-column matrix taking values 1's and 0's. Comparing the 3 predictions, if a student has only one possible outcome (i.e. only one 1, and two 0's) then the student is predicted to belong to the corresponding class. Otherwise, if there is uncertainty about the prediction because there is more than one 1 predicted for the student, then the class with the highest score is chosen to be the predicted one. 
\end{enumerate}
For instance, consider the following example: 
\begin{example}\label{example:multi-class}
	Suppose we have trained a classifier using 70\% of Dataset 1. When we infer the model on the test sample (remaining 30\%, consisting of 15 students), we obtain 3 vectors of scores, one for each class and we can compute their Gini Indices, see Figure \ref{Ex:avgGini}.
	
	\begin{figure}[t!]
		\centering
		\includegraphics[scale=0.6]{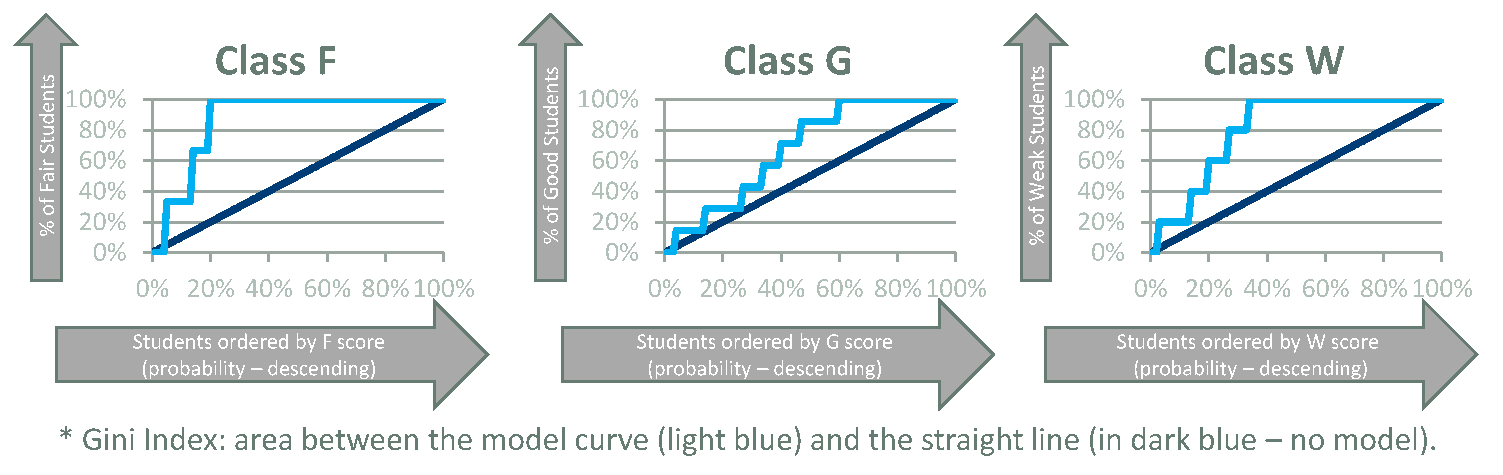}
		\caption{\label{Ex:avgGini}{Example - Averaged Gini Index Computation}}
	\end{figure}
	\begin{table}[t!]
	\centering
	\caption{Example - Predicting Classes}
	\scalebox{0.9}{%
		\begin{tabular}{|c|c|c|c|c|c|c|c|c|c|} 
			\hline
			\textbf{ID}&\textbf{Actual Class}&\textbf{score F}&\textbf{score G}&\textbf{score W}&\textbf{Max Pred.}&\textbf{F}&\textbf{G}&\textbf{W}&\textbf{Predicted Class}\\  \hline
			1 & F & 0.365 & 0.1 & 0.707 & W & 1 & 0 & 0 & F\\ \hline
			2 & F & 0.015 & 0.25 & 0.647 & W & 0 & 0 & 0 & W\\ \hline
			3 & G & 0.828 & 0.232 & 0.337 & F & 1 & 0 & 0 & F\\ \hline
			4 & G & 0.085 & 0.13 & 0.758 & W & 0 & 0 & 1 & W\\ \hline
			5 & W & 0.663 & 0.038 & 0.853 & W & 1 & 0 & 1 & W\\ \hline
			6 & G & 0.389 & 0.62 & 0.142 & G & 1 & 1 & 0 & G\\ \hline
			7 & G & 0.234 & 0.078 & 0.723 & W & 0 & 0 & 0 & W\\ \hline
			8 & W & 0 & 0.054 & 0.793 & W & 0 & 0 & 1 & W\\ \hline
			9 & W & 0.009 & 0 & 0.944 & W & 0 & 0 & 1 & W\\ \hline
			10 & G & 0.33 & 0.797 & 0 & G & 1 & 1 & 0 & G\\ \hline
			11 & F & 0.266 & 0.818 & 0.01 & G & 0 & 1 & 0 & G\\ \hline
			12 & G & 0.33 & 0.797 & 0 & G & 1 & 1 & 0 & G\\ \hline
			13 & G & 0.248 & 0.58 & 0.22 & G & 0 & 1 & 0 & G\\ \hline
			14 & W & 0.18 & 0.167 & 0.648 & W & 0 & 0 & 0 & W\\ \hline
			15 & W & 0.061 & 0.186 & 0.745 & W & 0 & 0 & 1 & W\\ \hline
	\end{tabular}}
	\label{tab:example_multi-class}
\end{table}
	In this example, the Gini Indices of Classes $F$, $G$, $W$ are 97.2\%, 76.8\%, 98\% respectively, hence the Averaged Gini Index is 90.7\%.\\
	We map the three scores linearly to the interval $[0,1]$, i.e. we normalize them to make them comparable. The normalized scores are represented in Table \ref{tab:example_multi-class} in columns \emph{score F}, \emph{score G}, \emph{score W}.\\ 
	Column \emph{Actual Class} corresponds to the actual target variable that we aim to predict. Treating each score as if it was the score associated to a binary classification problem, we need to set a threshold for each class such that if the score is greater than the threshold then the student belongs to such  class otherwise he/she doesn't (i.e., he/she belongs to one of the other two classes). Therefore we set three thresholds $\tau_F$, $\tau_G$, and $\tau_W$ for Class, $F$, $G$, and $W$ respectively. For instance, let $\tau_F=0.267$, $\tau_G=0.323$, and $\tau_W=0.740$. For student 1 in Table \ref{tab:example_multi-class}, the chance to be classified as $F$ is $0.365\geq \tau_F$, whereas the probabilities to belong to classes $G$ and $W$ are less than $\tau_G$ and $\tau_W$ respectively. In conclusion, once the three thresholds are set, we can claim that student 1 is a Fair student. \\
	Student 6 has score $F=0.389 \geq \tau_F$ and score $G=0.620\geq \tau_G$ so he/she belongs either to Class $F$ or to class $G$. Since the scores are normalized and are comparable, we set the predicted class to be the one corresponding to the highest score, hence we predict student ID=6 to belong to class $G$. \\
	For student 2 (7 and 14) note that the three scores are all below the thresholds so the predicted class is the one corresponding to the greatest score, i.e. the student is predicted as Weak.\\
	The max probability associated to each student is expressed in column \emph{Max Pred.}, and if we compare this column with column \emph{Actual Class} we note that taking the max score as the predicted class would not have been a good strategy. \\
	By setting the three thresholds $\tau_F$, $\tau_G$, and $\tau_W$ and considering the max score only in case of uncertainty we obtained for each student a predicted class, expressed in column \emph{Predicted Class}. If we compare the actual class with the predicted class we can build the corresponding confusion matrix:
	\begin{table}[h!]
		\centering
		\caption{Example - Confusion Matrix}
		\begin{tabular}{c|c c c}
			&           \textbf{F}  &  \textbf{G} &  \textbf{W}  \\
			\hline
			\textbf{F} &1 & 1 & 1\\
			\textbf{G} &1 & 4 & 2\\
			\textbf{W} & 0 & 0 & 5
		\end{tabular}	
	\end{table}
\end{example}
The threshold for class W in dataset 1 is typically higher than that for the other two classes due to the combination of two reasons. The first is that the test sample is fairly small. The second is that the number of class W instances is also small. As such, based on the fact that the threshold is determined by finding the score that results in the closest point on the ROC curve to the top left corner, the threshold has to be high in order to make sure that the points are identified correctly. Therefore, since the number of class W points is low, missing one of them would result in a significant drop in specificity and sensitivity. Thus, the optimal threshold should be high to be able to identify and classify them correctly.
\section{ML Parameter Tuning and Application} \label{ML_multi}
We chose one algorithm for each area of ML aiming to cover all types of classification methods including tree-based (RF), vector-based (SVM-RBF), distance-based (k-NN), regression-based (LR), probabilistic (NB), and neural network-based (NN1, NN2, and NN3 with 5 neurons per layer). The corresponding bagging ensemble models consist of all possible combinations of the aforementioned base models.
In Section \ref{multi-class_applications}, we explain how we train a NN. In the following sections, for each dataset, we show the impact of each variable on the performance of each classifier. As explained in Section \ref{sec:multi_intro}, in order to understand which variables are the best predictors for each algorithm, we decided to average, for each variable, its importance on every model and this average is assigned to the variable and defined to be its \emph{averaged importance}. In Section \ref{sec:multi_results} we will show that the most important variables affect the performances of some classifiers. 
\subsection{Neural Network Tuning}\label{multi-class_applications}
\begin{figure}[!h]
	\centering
	\includegraphics[scale=0.65]{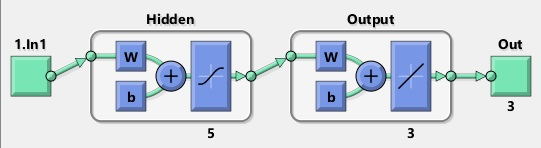}
	\caption{\label{DNN1}{NN with 1 hidden layer}}
\end{figure}

\begin{figure}[!h]
	\centering
	\includegraphics[scale=0.65]{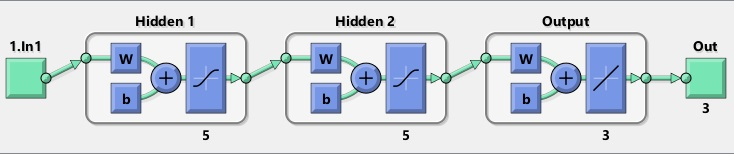}
	\caption{\label{DNN2}{NN with 2 hidden layers}}
\end{figure}

\begin{figure}[!h]
	\centering
	\includegraphics[scale=0.55]{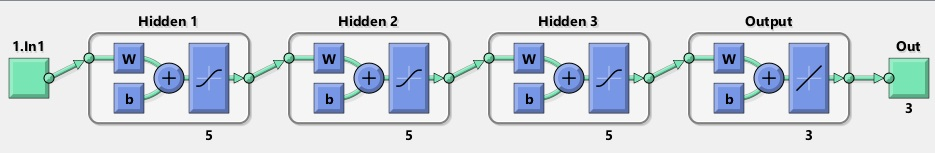}
	\caption{\label{DNN3}{NN with 3 hidden layers}}
\end{figure}

Finding the optimal number of neurons for NN is still an open field of research and requires a lot of computational resources. The authors in \cite{CH5_numb_layers} summarize some formulas for the computation of the optimal number of hidden neurons $N_h$:
\begin{itemize}
	\item $N_h = \frac{\sqrt{1+8 N_i}-1}{2}$
	\item $N_h = \sqrt{N_i N_o}$
	\item $N_h = \frac{4 N_i^2 + 3}{N_i^2 - 8}$
\end{itemize} 
where $N_i$ is the number of input neurons (number of variables) and $N_o$ is the number of output neurons (3 classes).
Applying the latter formulas to our datasets at the two different stages, we obtained a number of neurons between 2 and 6. Considering that we adopted the early stopping technique in order to prevent over-fitting and reduce variance, we decided to choose this number in the high range of the interval $[2,6]$ and set it to be equal to 5 instead of performing a full optimization (i.e., brute force searching). \\
The results obtained by using 1 hidden layer with 5 neurons were so promising that we decided to stress our hypothesis about early stopping and tried NN with 2 and 3 hidden layers with 5 neurons each, obtaining similar results. \\
The NN models we built are as in Figure \ref{DNN1}, \ref{DNN2}, \ref{DNN3}.

The initialization of the weights of neural networks was implemented by using the Nguyen-Widrow Initialization Method \cite{deep} whose goal is to speed up the training process by choosing the initial weights instead of generating them randomly. Simply put, this method assigns to each hidden node its own interval at the start of the training phase. By doing so, during the training each hidden layer has to adjust its interval size and location less than if the initial weights are chosen randomly. Consequently, the computational cost is reduced.\\
Levenberg-Marquardt backpropagation was used to train the models: this algorithm was introduced for the first time by Levenberg and Marquardt in \cite{deep4}, and is derived from Newton’s method that was designed for minimizing functions that are sums of squares of nonlinear functions \cite{deep3}. This method is confirmed to be the best choice in various learning scenarios, both in terms of time spent and performance achieved, \cite{deep5} .
Moreover, the datasets were normalized in input by mapping linearly to $[-1,1]$ (the activation function used in the input layer is the hyperbolic tangent) and in output to $[0,1]$ (the activation function in the output layer is linear) in order to avoid saturation of neurons and make the training smoother and faster.

\subsection{ML Algorithms' Parameter Tuning}
\indent\indent Hyper-parameter tuning has become an essential step to improve the performance of ML algorithms. This is due to the fact that each ML algorithm is governed by a set of parameters that dictate its predictive performance \cite{ee}. Several methods have been proposed in the literature to optimize and tune these parameters such as grid search algorithm, random search, evolutionary algorithms, and Bayesian optimization method \cite{ee,ff}. \\
\indent This work adopts the \emph{grid search} method to perform hyper-parameter tuning. Grid search optimization method is a well-known optimization method often used to hyper tune the parameters of ML classification techniques. Simply put, it discretizes the values for the set of techniques' parameters \cite{ee}. For every possible combination of parameters, the corresponding classification models are trained and assessed. Mathematically speaking, this can be formulated as follows:
\begin{equation}
\max\limits_{parm} f(parm)
\end{equation}
where $f$ is an objective function to be maximized (typically the accuracy of the model) and $parm$ is the set of parameters to be tuned.
Despite the fact that this may seem computationally heavy, grid search method benefits from the ability to perform the optimization in parallel, which results in a lower computational complexity \cite{ee}.\\
\indent In contrast to traditional hyper-parameter tuning algorithms that perform the optimization with the objective of maximizing the accuracy of the ML model, this work tunes the parameters used for each model using the \emph{grid search} optimization method to maximize the average Gini index (for more statistical significance and robustness \cite{gini_reason}) over multiple splits \cite{V}. More specifically, the objective function is:
\begin{equation}
\max\limits_{parm} Average\;Gini\;Index \; = \max\limits_{parm} \frac{1}{N}\sum\limits_{i=1}^{N} Gini\;Index_i(parm)
\end{equation}
where $parm$ is the set of parameters to be tuned for each ML algorithm and $N$ is the number of different splits considered. For example, in the case of K-NN algorithm, $parm=\{K\}$ which is the number of neighbors used to determine the class of the data point.\\
\indent R was used to implement the eight classifiers and the corresponding ensemble learners. As mentioned above, the eight classifiers considered in this work are SVM-RBF, LR, NB, k-NN, RF, NN1, NN2, and NN3. All the classifiers were trained using all the variables available. Moreover, the parameters of the algorithms were tuned by maximizing the Gini Index of each split. Furthermore, 200 different splits of data were used to reduce the bias of the models under consideration.\\ 
\begin{table}[!t]
	\centering
	\caption{Grid Search Parameter Tuning Range}
	\scalebox{0.9}{
		\begin{tabular}{|p{2cm}|p{4.5cm}|p{4.5cm}|} 
			\hline     
			\textbf{Algorithm} & \textbf{Parameter Range in Dataset 1}& \textbf{Parameter Range in Dataset 2}  \\
			\hline
			SVM-RBF & C=[0.25, 0.5, 1] \& sigma = [0.05-0.25] & C=[0.25, 0.5, 1] \& sigma = [0.5-3.5]  \\ \hline
			NB&usekernel=[True,False]&usekernel=[True,False]\\ \hline
			K-NN & k=[5,7,9,...,43]&k=[5,7,9,...,43] \\ \hline
			RF & mtry=[2,3,...,12]&mtry=[2,3,4]\\ \hline
	\end{tabular}}
	\label{tab:ML_model_parameter_range_ch5}
\end{table}
Table \ref{tab:ML_model_parameter_range_ch5} summarizes the range of values for the parameters of the different ML algorithms considered in this work. \\ 
\mbox{}\\  
\mbox{}\\
\mbox{}\\
Note the following:
\begin{itemize}
	\item For the NB algorithm, density estimator used by the algorithm is represented using the \emph{usekernel} parameter. In particular,  \emph{usekernel=false} means that the data distribution is assumed to be Gaussian. On the other hand, \emph{usekernel = true} means that the data distribution is assumed to be non-Gaussian.
	\item The LR algorithm was not included in the table. This is due to the fact that it has no parameters to optimize. The sigmoid function, which is the default function, was used by the grid search method to maximize the Gini index.
	\item The NN method was not included in the table because it was explained in the previous Section \ref{multi-class_applications}.
\end{itemize}
The features are ordered according to their importance. This is done for the two datasets and for each of the algorithm used. This provides us with better insights about which features are important for each algorithm and each dataset. 
The importance of the features is determined using the CARET package available for R language \cite{feature_importance_caret}. Depending on the classification model adopted, the importance is calculated in one of multiple ways. For example, when using RF method, the prediction accuracy on the out-of-bag portion of the data is recorded. This is iteratively done after permuting each predictor variable. The difference between the two accuracy values is then averaged over all trees and normalized by the standard error \cite{feature_importance_caret}. In contrast, when the k-NN method is used, the difference between the class centroid and the overall centroid is used to measure the variable influence. Accordingly, the separation between the classes is larger whenever the difference between the class centroids is larger \cite{feature_importance_caret}. On the other hand, when using the NN method, the CARET package uses the same feature importance method proposed in Gevrey \textit{et al.} which uses combinations of the absolute values of the weights \cite{feature_importance_nn}. This importance is reflected in the weights calculated for each feature for each classification model with more important features contributing more towards the prediction. \\ 
The final step consist of selecting the most suitable bagging ensemble learner for both datasets at the two course delivery stages. 
\subsection{Features importance: Dataset 1 - Stage 20\%}
\begin{itemize}
	\item RF:
	The variables' importance in terms of predictivity is described in Table \ref{tab:table_dataset_1_vars_weight_multi} that shows that the most relevant features are ES2.2 and ES3.3.
	\begin{table}[!t]
		\centering
		\caption{ Dataset 1 - Stage 20\% - Features' importance for Different Base Classifiers}
		\scalebox{0.9}{%
			\begin{tabular}{|c|c|c|c|c|c|c|c|c|} 
				\hline     
				\textbf{Ranking} & \textbf{RF} & \textbf{SVM-RBF}& \textbf{NN1}& \textbf{NN2}& \textbf{NN3}& \textbf{k-NN}& \textbf{LR}& \textbf{NB} \\
				\hline
				1 & ES2.2 &ES2.2 &ES2.2 &ES2.2 &ES2.2 &ES2.2 &ES1.1 &ES2.2  \\
				\hline
				2 & ES3.3 &ES3.3 &ES3.5 &ES3.3 &ES3.3 &ES3.3 &ES1.2 &ES3.3 \\
				\hline
				3 & ES2.1 &ES2.1 &ES3.3 &ES3.5 &ES3.5 &ES2.1 &ES3.5 &ES2.1 \\
				\hline
				4 & ES1.1 &ES3.5 &ES3.2 &ES2.1 &ES2.1 &ES3.5 &ES3.3 &ES3.5 \\
				\hline
				5 & ES3.5 &ES1.2 &ES1.1 &ES3.2 &ES3.2 &ES3.4 &ES3.4 &ES3.4 \\
				\hline
				6 & ES1.2 &ES3.4 &ES2.1 &ES1.1 &ES1.1 &ES1.2 &ES3.2 &ES1.2 \\
				\hline
				7 & ES3.4 &ES1.1 &ES1.2 &ES3.4 &ES3.4 &ES1.1 &ES3.1 &ES1.1 \\
				\hline
				8 & ES3.1 &ES3.1 &ES3.4 &ES1.2 &ES1.2 &ES3.1 &ES2.2 &ES3.2 \\
				\hline
				9 & ES3.2 &ES3.2 &ES3.1 &ES3.1 &ES3.1 &ES3.2 &ES2.1 &ES3.1 \\
				\hline
		\end{tabular}}
		\label{tab:table_dataset_1_vars_weight_multi}
	\end{table}
	\item SVM-RBF:
	The variables' importance for SVM is described in Table \ref{tab:table_dataset_1_vars_weight_multi}, that shows that the most relevant features are ES2.2 and ES3.3.
	\item NN1: 
	For NN1, the variables' importance in terms of predicativity is described in Table \ref{tab:table_dataset_1_vars_weight_multi} that shows that the most relevant features are ES2.2 and ES3.5.
	\item NN2: 
	The most important variables for NN2 are ES2.2 and ES3.2, as shown in Table \ref{tab:table_dataset_1_vars_weight_multi}.
	\item NN3:  
	The variables' importance in terms of predicativity is described in Table \ref{tab:table_dataset_1_vars_weight_multi} that shows that the most relevant features are ES2.2 and ES3.2.
	\item k-NN:
	Table \ref{tab:table_dataset_1_vars_weight_multi} shows that the most relevant features for k-NN are ES2.2 and ES3.3.
%
	\item LR:
	The variables' importance in terms of predicativity is described in Table \ref{tab:table_dataset_1_vars_weight_multi} that shows that the most relevant features are ES1.1 and ES1.2.
	\item NB: Table \ref{tab:table_dataset_1_vars_weight_multi} shows that the most relevant features are ES2.2 and ES3.3.
\end{itemize}
\subsection{Features importance: Dataset 1 - Stage 50\%}
It is important to point out that, for Dataset 1 at stage 50\%, features ES4.1 and ES4.2 are the most important for every classifier. 
\begin{itemize}
	\item RF: For RF, the variables' importance in terms of predicativity is described in Table \ref{tab:table_dataset_1_50_predictors_multi} that shows that the most relevant features are ES4.1 and ES4.2.	
	\item SVM-RBF: The variables' importance in terms of predicativity is described in Table \ref{tab:table_dataset_1_50_predictors_multi} that shows that the most relevant features are ES4.1 and ES4.2.	
%
	\item NN1:
	The variables' importance in terms of predicativity is described in Table \ref{tab:table_dataset_1_50_predictors_multi} that shows that the most relevant features are ES4.1 and ES4.2.	
%
%
	\item NN2:
	The variables' importance in terms of predicativity is described in Table \ref{tab:table_dataset_1_50_predictors_multi} that shows that the most relevant features are ES4.1 and ES4.2.	
%
%
	\item NN3: 
	The variables' importance in terms of predicativity is described in Table \ref{tab:table_dataset_1_50_predictors_multi} that shows that the most relevant features are ES4.1 and ES4.2.	
%
%
	\item k-NN: Table \ref{tab:table_dataset_1_50_predictors_multi} shows that the most relevant features for k-NN are ES4.1 and ES4.2.	
%
%
	\item LR: Table \ref{tab:table_dataset_1_50_predictors_multi} shows that the most relevant features for LR are ES4.1 and ES4.2.	
%
%
	\item NB: The variables' importance in terms of predicativity is described in Table \ref{tab:table_dataset_1_50_predictors_multi} that shows that the most relevant features are ES4.1 and ES4.2.	
%
%
\end{itemize}
	\begin{table}[!h]
	\centering
	\caption{Dataset 1 - Stage 50\% -  Features' importance for Different Base Classifiers}
	\scalebox{0.9}{%
		\begin{tabular}{|c|c|c|c|c|c|c|c|c|} 
			\hline     
			\textbf{Ranking} & \textbf{RF} & \textbf{SVM-RBF}& \textbf{NN1}& \textbf{NN2}& \textbf{NN3}& \textbf{k-NN}& \textbf{LR}& \textbf{NB} \\
			\hline
			1 & ES4.1 &ES4.1 &ES4.1 &ES4.1 &ES4.1 &ES4.1 &ES4.1 &ES4.1 \\
			\hline
			2 & ES4.2 &ES4.2 &ES4.2 &ES4.2 &ES4.2 &ES4.2 &ES4.2 &ES4.2 \\
			\hline
			3 & ES2.2 &ES2.2 &ES3.3 &ES3.5 &ES5.1 &ES2.2 &ES1.1 &ES1.1 \\
			\hline
			4 & ES5.1 &ES5.1 &ES3.5 &ES3.3 &ES3.5 &ES5.1 &ES2.1 &ES2.1 \\
			\hline
			5 & ES2.1 &ES3.3 &ES2.1 &ES5.1 &ES3.3 &ES3.3 &ES1.2 &ES1.2 \\
			\hline
			6 & ES1.1 &ES2.1 &ES5.1 &ES2.1 &ES2.1 &ES2.1 &ES3.3 &ES3.3\\
			\hline
			7 & ES3.5 &ES3.5 &ES1.1 &ES3.4 &ES2.2 &ES3.5 &ES3.4 &ES3.4 \\
			\hline
			8 & ES3.3 &ES1.2 &ES3.4 &ES2.2 &ES1.1 &ES3.4 &ES5.1 &ES5.1 \\
			\hline
			9 & ES3.4 &ES3.4 &ES3.2 &ES1.1 &ES3.4 &ES1.2 &ES3.5 &ES3.5 \\
			\hline
			10 & ES3.1 &ES1.1 &ES2.2 &ES3.1 &ES3.2 &ES1.1 &ES2.2 &ES2.2 \\
			\hline
			11 & ES3.2 &ES3.1 &ES1.2 &ES3.1 &ES3.1 &ES3.1 &ES3.1 &ES3.1 \\
			\hline
			12 & ES1.2 &ES3.2 &ES3.1 &ES1.2 &ES1.2 &ES3.2 &ES3.2 &ES3.2 \\
			\hline
	\end{tabular}}
	\label{tab:table_dataset_1_50_predictors_multi}
\end{table}
In general, the most important features for almost all the classifiers are ES4.1 and ES4.2. These features correspond to the \emph{Evaluate} category as per Bloom's taxonomy which represents one of the highest level of comprehension of the course material from the educational point of view. Therefore, it makes sense for these features to be suitable indicators and predictors of student performance. 
\begin{table}[!h]
	\centering
	\caption{Dataset 2 - Stage 20\% - Features' importance}
	\scalebox{0.9}{
		\begin{tabular}{|c|c|} 
			\hline     
			\textbf{Ranking} & \textbf{Feature}  \\
			\hline
			1 &Assignment01   \\
			\hline
			2 & Quiz01   \\
			\hline
	\end{tabular}}
	\label{tab:table_dataset_2_20_predictors_multi}
\end{table}
\subsection{Features importance: Dataset 2 - Stage 20\%}
We have only two features for Dataset 2, Stage 20\% and for all the classifiers, the list of features ordered by importance, see Table \ref{tab:table_dataset_2_20_predictors_multi}. \\
Since Dataset 2 at stage 20\% has only two variables we can represent it graphically in order to have a better understanding of the situation and to explain why all the algorithms agree that Assignment01 is the most important predictor. \\
\begin{figure}[!h]
	\centering
	\includegraphics[scale=0.35]{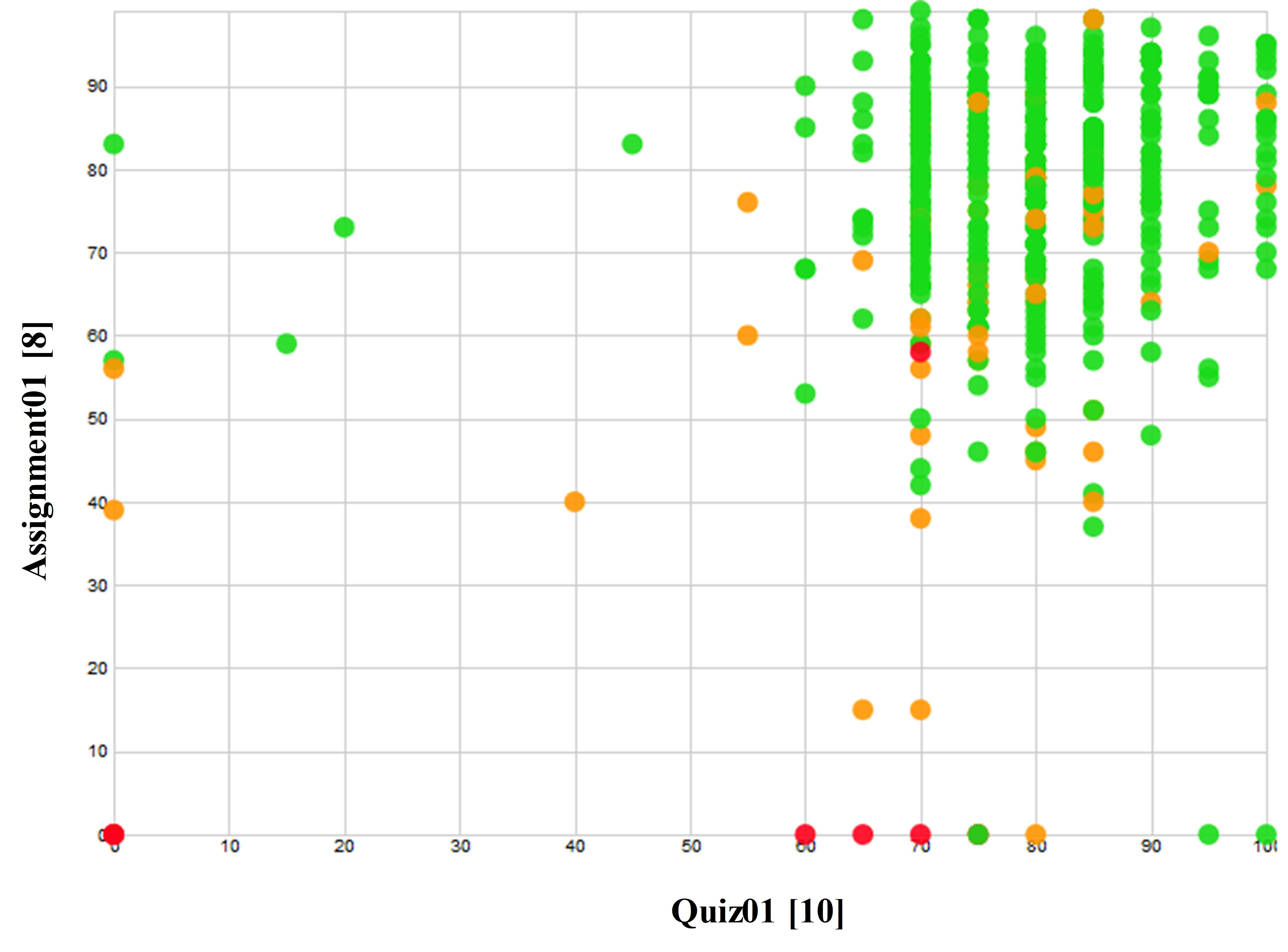}
	\caption{\label{big20_multi_plot}{Dataset 2 - Stage 20\% - scatter plot}}
\end{figure}
Figure \ref{big20_multi_plot} shows that it is straightforward to identify the categories of students by setting some thresholds on the Assignment01 feature. For instance, most of the Weak students have grade zero in Assignment01.
\subsection{Features importance: Dataset 2 - Stage 50\%}
The variables' importance for NN1, NN2, k-NN, and NB is described in Table \ref{tab:table_dataset_2_50_vars_weight_multi_A}, whereas the variables' importance for NN3, LR, RF, and SVM is described in Table \ref{tab:table_dataset_2_50_vars_weight_multi_B}.
\begin{table}[!h]
	\centering
	\caption{Dataset 2 - Stage 50\% - NN1, NN2, k-NN, and NB, Features' importance}
	\scalebox{0.9}{
		\begin{tabular}{|c|c|} 
			\hline     
			\textbf{Ranking} & \textbf{Feature} \\
			\hline
			1 &Assignment02   \\
			\hline
			2 & Assignment01  \\
			\hline
			3 & Midterm Exam   \\
			\hline
			4 &  Quiz01   \\
			\hline
	\end{tabular}}
	\label{tab:table_dataset_2_50_vars_weight_multi_A}
\end{table}
\begin{table}[!h]
	\centering
	\caption{Dataset 2 - Stage 50\% - NN3, LR, RF, and SVM-RBF, Features' importance}
	\scalebox{0.9}{
		\begin{tabular}{|c|c|} 
			\hline     
			\textbf{Ranking} & \textbf{Feature} \\
			\hline
			1 & Assignment01  \\
			\hline
			2 & Assignment02   \\
			\hline
			3 & Midterm Exam    \\
			\hline
			4 &  Quiz01   \\
			\hline
	\end{tabular}}
	\label{tab:table_dataset_2_50_vars_weight_multi_B}
\end{table}

Based on the aforementioned results, it  can be seen that assignments are better indicators of the student performance. This can be attributed to several factors. The first is the fact that assignments typically allow instructors to assess the three higher levels of cognition as per Bloom's taxonomy, namely analysis, synthesis, and evaluation \cite{assignment_importance}. As such, assignments provide a better indicator of the learning level that a student has achieved and consequently can give insights about his/her potential performance in the class overall. Another factor is that students tend to have more time to complete assignments. Moreover, they are often allowed to discuss issues and problems among themselves. Thus, students not performing well in the assignments may be indicative of them not fully comprehending the material. This can result in the students receiving a lower overall final course grade.
\section{Experimental Results and Discussion}\label{sec:multi_results}
Matlab 2018 was used to build the Neural Networks classifiers, whereas all the other models were built using R.\\
All possible combinations of ensembles of eight baggings of models (256 in total) were computed for the initial Train-Test split and for 5 extra splits. 
For each dataset, the average of the performances, namely \emph{averaged Gini Index}, on the 6 splits was used to select the most robust ensemble learner. In addition, we computed the p-values of all the ensembles for all the splits aiming to select the ensemble learner \emph{with highest averaged Gini index that was also statistically significant on every split}. Note that the contribution of each feature is determined by the base learner model being used in the ensemble as per the ranking determined for each dataset at each stage. For example, if the RF learner is part of the ensemble being considered for Dataset 1 at 50\% stage, the first split is done over feature ES 4.1, the second split is over feature ES 4.2, and so on.\\ 
In the following sections we will see the results obtained for the two datasets at each stage.
\subsection{Results: Dataset 1 - Stage 20\%}
\begin{figure}[!htbp]
	\centering
	\includegraphics[scale=0.75]{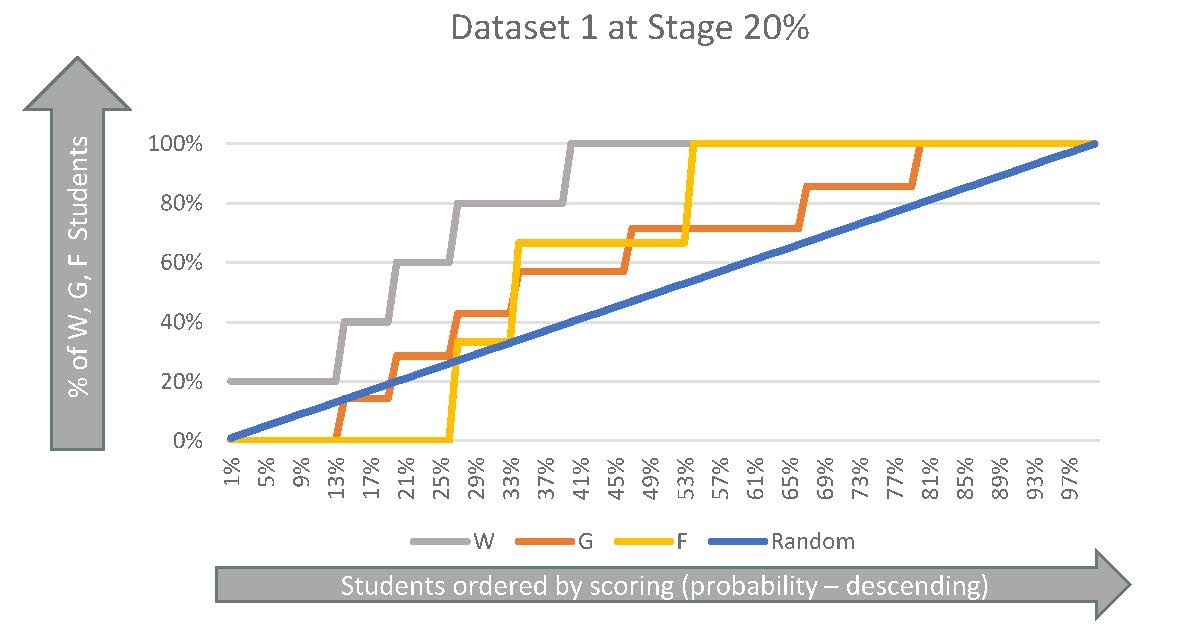}
	\caption{\label{Performance_Ensemble_GFW_small20}{Dataset 1 - Stage 20\% - Ensemble Learner}}
\end{figure}

If we based our choice only on the Gini index  corresponding to the initial split, the ensemble learner we would have selected for Dataset 1 at Stage 20\% would have been formed by NB, NN1, and SVM-RBF.
Instead, the ensemble learner that appears to be the most stable on every split and with statistical significance is the one formed by a bagging of the NN2 model and the combination of the bagging of NN2 and NB as a bagging ensemble. Figure \ref{Performance_Ensemble_GFW_small20} shows the results obtained by inferring the ensemble on the initial test sample. 

Classes G, F, W have Gini Indices equal to 46.4\%, 38.9\% and 94.0\% respectively. Hence, the Averaged Gini Index is 59.8\%. On average, on Test sample and the 5 extra splits the Averaged Gini Index is 62.1\%.  The corresponding p-values are all less than 0.03.\\
The confusion matrix for the Test sample (consisting of 15 students), obtained as explained in Section \ref{sec.method}, is shown in Table \ref{data1_20_confusion}. 

\begin{table}[t!]
	\centering
	\caption{Dataset 1 - Stage 20\% Ensemble (NN2) Confusion Matrix \\  $\tau_F=0.158, \tau_G=0.310,\tau_W=0.682$}
	\begin{tabular}{c|c c c}
		&           \textbf{F}  &  \textbf{G} &  \textbf{W}  \\
		\hline
		\textbf{F} &1 & 1 & 1\\
		\textbf{G} & 1 & 4 & 2\\
		\textbf{W} & 0 & 0 & 5
	\end{tabular}
	\label{data1_20_confusion}
\end{table}

Table \ref{tab:table_dataset_1_20multiperf} illustrates the performances of the ensemble learner in terms of precision, recall, F-measure and false positive rate per class and on average. These quantities depend on the thresholds $\tau_F$, $\tau_G$ and $\tau_W$ and the way we defined the predictions. The Accuracy is 66.7\%. Although this may seem to be low, it actually outperforms all of the base learners used to create the bagging ensemble. Note that the low accuracy may be attributed to the fact that the dataset itself is small and hence did not have enough instances to learn from. 

\begin{table}[h!]
	\centering
	\caption{Dataset 1 - Stage 20\% - Ensemble Performances}
	\scalebox{1}{
		\begin{tabular}{|c|c c c c|} 
			\hline
			&\textbf{Precision}&\textbf{Recall}&\textbf{F-measure}&\textbf{False Positive Rate}\\  \hline
			\textbf{F}    &0.33&0.50&0.40&0.50 \\       
			\textbf{G}    &0.57&0.80&0.67&0.20 \\     
			\textbf{W}   &1.00&0.63&0.77&0.38 \\       \hline
			\textbf{Avg}&0.64&0.64&0.61&0.36 \\       \hline
	\end{tabular}}
	
	\label{tab:table_dataset_1_20multiperf}
	
\end{table}

\subsection{Results: Dataset 1 - Stage 50\%}
\begin{figure}[!htb]
	\centering
	\includegraphics[scale=0.75]{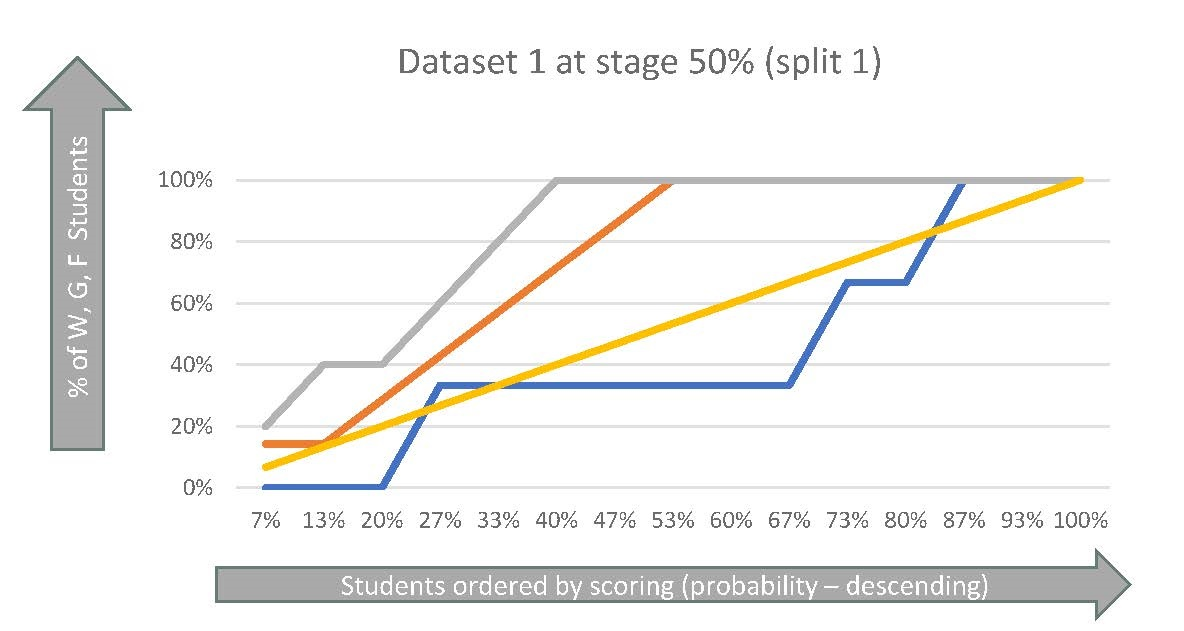}
	\caption{\label{Performance_Ensemble_GFW_small50}{Dataset 1 - Stage 50\% - Ensemble Learner}}
\end{figure}

For Dataset 1 at Stage 50\%, \emph{none of the ensembles we constructed were statistically significant} even if their Averaged Gini Indices are on average higher than the ones obtained for Dataset 1 at Stage 20\%. In fact, the performance for class F gets worse when we add the three variables. More precisely, when we add Features \emph{ES4.1}, \emph{ES4.2} and \emph{ES5.1} to Dataset 1 at stage 20\% obtaining  Dataset 1 at stage 50\%, they end up being the ones that have the main impact on the predictions. These variables help distinguish between W and G and in fact the performance corresponding to these two classes improve. However, since Fair students are closely correlated with the Good students class, the classifier becomes less confident in predicting the Fair students. \\    
The best ensemble in terms of performance is the one obtained from a bagging of NB and k-NN. The Averaged Gini Index on 6 splits is 74.9\% and on the initial test sample the Averaged Gini Index is 86.5\%. Figure \ref{Performance_Ensemble_GFW_small50} shows the performance obtained on Split 1, having Averaged Gini Index equals 50\%, with Gini Indices -22.2\%, 76.8\%, 86.0\% respectively on Classes F, G and W. On a different split, the ensemble formed by a bagging of NB and k-NN on Dataset 1 at stage 20\% gives Gini Indices 77.8\%, 53.6\% and 48.0\% respectively on Classes F, G and W. This proves that the performance heavily depends on the split. In general, when we add the new three features (obtaining Dataset1 at stage 50\%), the performance improves on classes G and W whereas it gets much worse for class F. \\
\mbox{}\\
\mbox{}\\
The confusion matrix obtained is the following: 
\begin{table}[!h]
	\centering
	\caption{Dataset 1 - Stage 50\% Ensemble (NB and k-NN) Confusion Matrix \\  $\tau_F=0.10, \tau_G=0.29,\tau_W=0.88$ }
	\begin{tabular}{c|c c c}
		&           \textbf{F}  &  \textbf{G} &  \textbf{W}  \\
		\hline
		\textbf{F}   & 0 & 2 & 1\\
		\textbf{G}   & 0 & 7 & 0\\
		\textbf{W}  & 1 & 1 & 3
	\end{tabular}
	
\end{table}

Table \ref{tab:table_dataset_1_50multiperf} illustrates the performances of the ensemble learner in terms of precision, recall, F-measure and false positive rate per class and on average. These quantities depend on the thresholds $\tau_F$, $\tau_G$ and $\tau_W$ and the way we defined the predictions. The Accuracy is 66.7\%. Again, the bagging ensemble outperforms all of the base learners used to create it despite it potentially being low. This is due to the fact that the dataset itself is small and hence did not have enough instances for the ensemble to learn from.
Note that we cannot compute the F-measure for class F as Precision and Recall are zero.
\begin{table}[!h]
	\centering
	\caption{Dataset 1 - Stage 50\% - Ensemble Performances}
	\scalebox{1}{
		\begin{tabular}{|c|c c c c|} 
			\hline
			&\textbf{Precision}&\textbf{Recall}&\textbf{F-measure}&\textbf{False Positive Rate}\\  \hline
			\textbf{F}    &0.00&0.00& - &1.00 \\       
			\textbf{G}    &1.00&0.70&0.82&0.30 \\     
			\textbf{W}   &0.60&0.75&0.67&0.25 \\       \hline
			\textbf{Avg}&0.53&0.48& - &0.52 \\       \hline
	\end{tabular}}
	\label{tab:table_dataset_1_50multiperf}	
\end{table}
It is worth noting that the low average Gini index can be attributed to 2 main reasons: 
\begin{itemize}
	\item This dataset is a small dataset.
	\item The Fair class is highly correlated with the Good students class. Hence, this is causing some confusion to the models being trained.
\end{itemize} 
This is further highlighted by the large false positive rate obtained for the Fair class. 

\subsection{Results: Dataset 2 - Stage 20\%}
\begin{figure}[!h]
	\centering
	\includegraphics[scale=0.75]{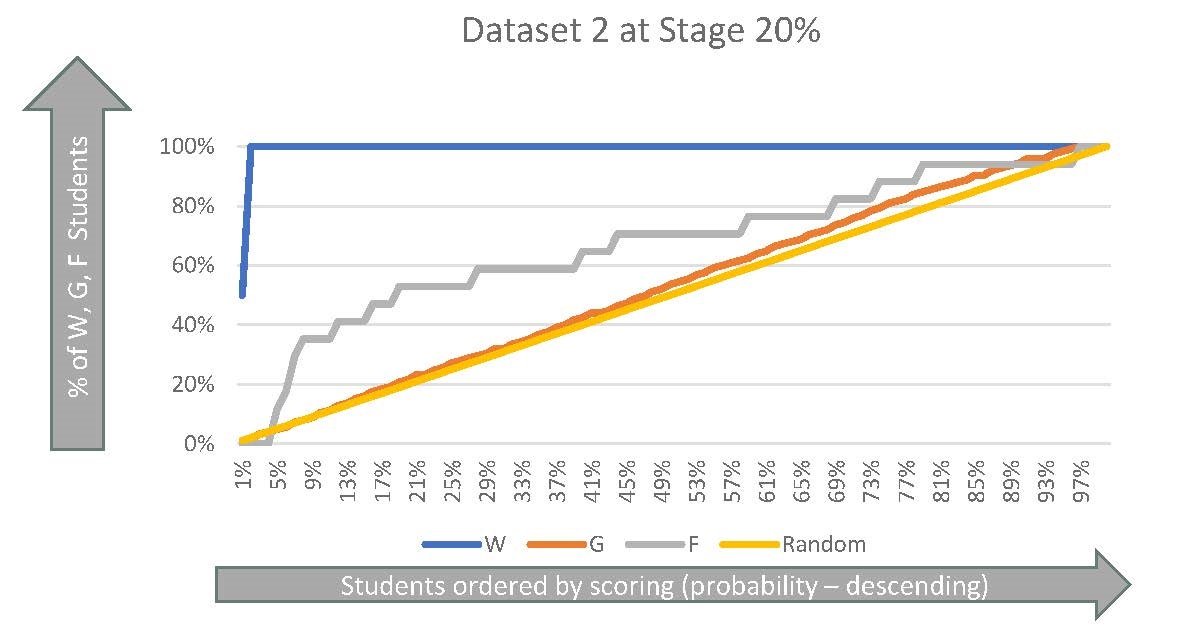}
	\caption{\label{Performance_Ensemble_GFW_big20}{Dataset 2 - Stage 20\% - Ensemble Learner}}
\end{figure}

The ensemble learner selected for Dataset 2 at Stage 20\% is formed by bagging of NB, k-NN, LR, NN2, and SVM-RBF. For instance, we show the results corresponding to the initial test sample. For each class, we normalized the scores obtained by the five baggings of models on the test sample in order to make these probabilities comparable, then we averaged them. The performances obtained are shown in Figure \ref{Performance_Ensemble_GFW_big20}.

Classes G, F, W have Gini Indices equal to 48.1\%, 38.6\% and 99.7\% respectively. The confusion matrix associated is shown in Table \ref{tab:table_dataset_2_confusion}.
\begin{table}[!h]
	\centering
	\caption{ Dataset 2 - Stage 20\% Ensemble (NB,k-NN,LR,NN2,SVM) Confusion Matrix \\  $\tau_F=0.12, \tau_G=0.62,\tau_W=0.40$}
	\begin{tabular}{c|c c c}
		&           \textbf{F}  &  \textbf{G} &  \textbf{W}  \\
		\hline
		\textbf{F} & 5 & 11 & 1\\
		\textbf{G} & 5 & 120 & 0\\
		\textbf{W} & 0 & 0 & 2
	\end{tabular}
	\label{tab:table_dataset_2_confusion}	
\end{table}

Furthermore, Table \ref{tab:table_dataset_2_20multiperf} illustrates the performances of the ensemble learner in terms of precision, recall, F-measure and false positive rate per class and on average. These quantities depend on the thresholds $\tau_F$, $\tau_G$ and $\tau_W$ and the way we defined the predictions. The Accuracy is 88.2\%, which is very good compared with respect to the performances obtained for Dataset 1. In a similar fashion to Dataset 1, the bagging ensemble outperforms the base learners in terms of classification accuracy. 

\begin{table}[!htb]
	\centering
	\caption{Dataset 2 - Stage 20\%  - Ensemble Performances}
	\scalebox{1}{
		\begin{tabular}{|c|c c c c|}
			\hline
			&\textbf{Precision}&\textbf{Recall}&\textbf{F-measure}&\textbf{False Positive Rate}\\  \hline
			\textbf{F}    &0.29&0.50&0.37&0.50 \\       
			\textbf{G}    &0.96&0.92&0.94&0.08 \\     
			\textbf{W}   &1.00&0.67&0.80&0.33 \\       \hline
			\textbf{Avg}&0.75&0.69&0.70&0.31 \\       \hline
	\end{tabular}}
	\label{tab:table_dataset_2_20multiperf}	
\end{table}

\subsection{Results: Dataset 2 - Stage 50\%}
\begin{figure}[!htb]
	\centering
	\includegraphics[scale=0.75]{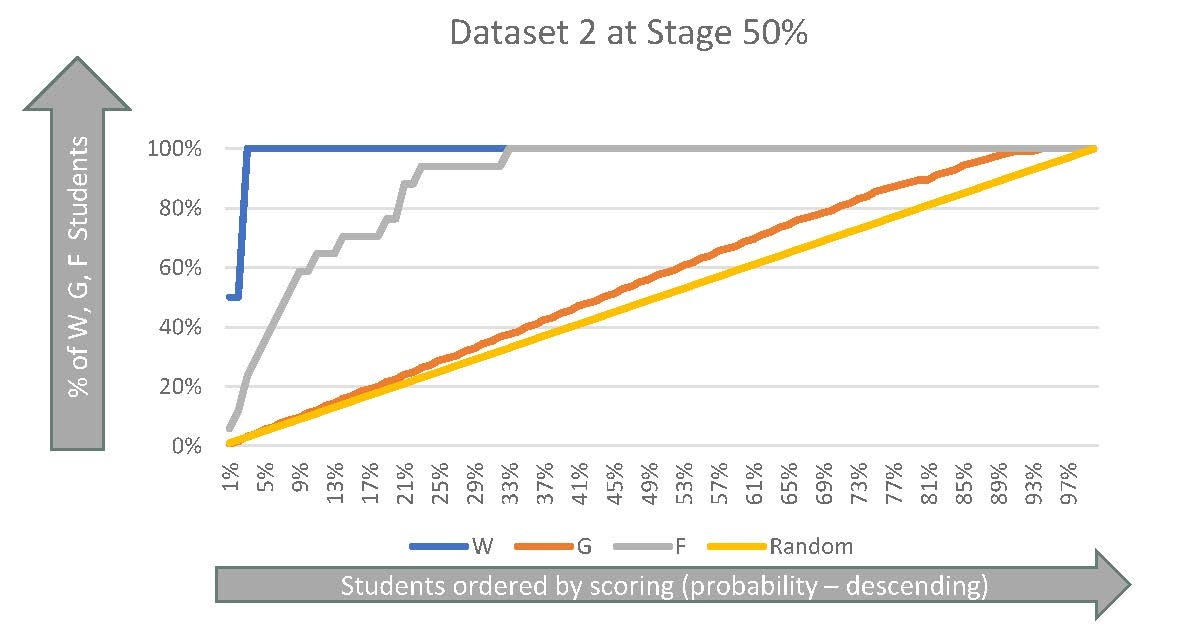}
	\caption{\label{Performance_Ensemble_GFW_uwo50}{Dataset 2 - Stage 50\% - Ensemble Learner}}
\end{figure}

The ensemble learner selected for Dataset 2 at Stage 50\% is formed by a bagging of LR models only.
The performances obtained on the initial test sample are shown in Figure \ref{Performance_Ensemble_GFW_uwo50}. On average, almost all the ensembles we constructed have very good performances and are statistically significant. The ensemble we selected is very robust on every split.

Classes G, F, W have Gini Indices equal to 92.3\%, 90.7\% and 99.3\% respectively. \\
The confusion matrix obtained is the following: 
\begin{table}[!tb]
	\centering
	\caption{Dataset 2 - stage 50\% Ensemble (LR) Confusion Matrix \\  $\tau_F=0.12, \tau_G=0.62,\tau_W=0.30$}
	\begin{tabular}{c|c c c}
		&           \textbf{F}  &  \textbf{G} &  \textbf{W}  \\
		\hline
		\textbf{F}   & 11 & 5 & 1\\
		\textbf{G}   & 3 & 122 & 0\\
		\textbf{W}  & 1 & 0 & 1
	\end{tabular}
	
\end{table}

Table \ref{tab:table_dataset_2_50multiperf} illustrates the performances of the ensemble learner in terms of precision, recall, F-measure and false positive rate per class and on average. These quantities depend on the thresholds $\tau_F$, $\tau_G$ and $\tau_W$ and the way we defined the predictions. The Accuracy is 93.1\%. Again, the bagging ensemble at this stage also outperforms the base learners in terms of classification accuracy. 
\begin{table}[!t]
	\centering
	\caption{Dataset 2 - Stage 50\% - Ensemble Performances}
	\scalebox{1}{
		\begin{tabular}{|c|c c c c|} 
			\hline
			&\textbf{Precision}&\textbf{Recall}&\textbf{F-measure}&\textbf{False Positive Rate}\\  \hline
			\textbf{F}    &0.65&0.73&0.69&0.27 \\       
			\textbf{G}    &0.98&0.96&0.97&0.04 \\     
			\textbf{W}   &0.50&0.50&0.50&0.50 \\       \hline
			\textbf{Avg}&0.71&0.73&0.72&0.27 \\       \hline
	\end{tabular}}
	\label{tab:table_dataset_2_50multiperf}
\end{table}

\subsection{Performance Comparison With Base Learners}
\indent Table \ref{perf_mat_base} shows the classification accuracy of the different base learners in comparison with the average accuracy across the 256 splits of the bagging ensemble. It can be seen that the bagging ensemble on average outperforms all of the base learners at the two course delivery stages for both datasets. This is despite the fact that some of the splits may have had poor distribution which often leads to lower classification accuracy of the ensemble. This further highlights and emphasizes the effectiveness of the proposed ensemble in accurately predicting and identifying students who may need help.
\begin{table*}[!h]
	\centering
	\caption{Performance of Bagging Ensemble and Base Learners\label{perf_mat_base}}
	\scalebox{1}{%
		\begin{tabular}{|l|c|c|c|c|}
			\hline
			\multicolumn{5}{|c|}{\textbf{Accuracy}}\\  \hline
			\textbf{Technique}&\multicolumn{2}{c|}{\textbf{Dataset 1}}&\multicolumn{2}{c|}{\textbf{Dataset 2}}\\ \hline
			&Stage 20\%&Stage 50\%&Stage 20\%&Stage 50\%\\ \hline
			RF&46.7\%&66.6\%&82.8\%&89.0\%\\ \hline
			NN&66.7\%&60\%&86.2\%&91.7\%\\ \hline
			K-NN&60\%&66.6\%&86.2\%&89.0\%\\ \hline
			NB&53.3\%&66.6\%&85.5\%&85.5\%\\ \hline
			LR&53.3\%&53.3\%&86.9\%&90.3\%\\ \hline
			SVM&46.7\%&33.3\%&86.2\%&90.3\%\\ \hline
			Ensemble&\textbf{66.7\%}&\textbf{66.7\%}&\textbf{88.2\%}&\textbf{93.1\%}\\ \hline
	\end{tabular}}
\end{table*}
\subsection{Results Summary}
The performances obtained for Dataset 1 and Dataset 2 are very different.
For Dataset 1, the models performances depend strongly on the splits. For instance, the same ensemble might perform very well on certain splits but have very low Averaged Gini Index on others, due to a negative Gini index on class F.
Moreover, only 25\% of the ensembles for Dataset 1 at the 20\% had averaged Gini Index above 50\% and of all the ensembles only one of them is statistically significant, the one corresponding to a bagging of NN2 models.\\
Although the evidence shows that this ensemble performs decently on each split we have considered for our experiments, we cannot assume that this is true on every other possible split we might have chosen instead. The problem is so dependent on the split selected, that even the ensemble we chose results in lack of robustness and poor performances.\\
For Dataset 1 at stage 50\%, the averaged Gini Index is in general higher than the averaged Gini Index obtained at stage 20\% because the Gini Indices corresponding to classes G (Good students) and class W (Weak students) improve when we add the three features \emph{ES4.1, ES4.2, ES5.1}. 
Since the Fair students class is highly correlated with the Good students class, the consequence is that when we add the best predictors, they predict incorrectly the Fair students. Consequently, the Gini Index for class F for each ensemble and for almost every split is negative or very low, leading to statistically insignificant results. In particular, there is not an ensemble among the 256 constructed such that the p-value corresponding to class F is lower than $0.03$ on every split. Note that the ensemble chosen at the 50\% stage is bagging of NB and k-NN. Although the ensemble was not statistically significant due to class F, it was statistically significant for the target class W.\\
For this reason, even though for completeness we are going to show the results for Dataset 1 at both stages, it is important to point out that if we were aiming to classify correctly the students for Dataset 1 and to use the classifier for applications in real world, we should not include the last three features, i.e. we should use Dataset 1 at stage 20\%.\\
Dataset 2 was easier to deal with and also the choice of the best ensemble was straightforward. 
For Dataset 2 at stage 50\%, 88\% of the ensembles have averaged Gini Indices above 90\%, and 96\% of the ensembles were statistically significant. \\
For Dataset 2, the highest averaged Gini Index led us to choose:
\begin{itemize}
	\item the ensemble of baggings of NB, k-NN, LR, and NN2 for the 20\% stage. 
	\item the ensemble consisting of bagging of LR for the 50\% stage.
\end{itemize}
Note that in general, it is better to perform the prediction at the 50\% stage rather than at the 20\% stage. This is due to the fact that more features are collected at the 50\% stage, resulting in the learners being able to gain more information. Although this observation was not evident for dataset 1, this is due to the dataset being small with only a few instances of the F class that were at the border between the G and W classes. However, it was observed that the F-measure was high at both stages for the target class W.\\
For dataset 2, the results showed that indeed predicting at the 50\% stage is better since the performance of the ensemble improved with the added number of features. However, the results at the 20\% stage were still valuable as they helped provide vital insights at an extremely early stage of the course delivery as evident by the fact that the F-measure was close to 0.7 at that stage.
\section{Conclusion, Research Limitations, and Future Work} \label{sec:future2}
In this paper, we investigated the problem of identifying student who may need help during course delivery time for an e-Learning environment. The goal was to predict the students' performance by classifying them into one of three possible classes, namely Good, Fair, and Weak. In particular, we tackled this multi-class classification problem for two educational datasets at two different course delivery stages, namely at the 20\% and 50\% mark. We trained eight baggings of models for each dataset and considered all the possible ensembles that could be generated by considering the scores produced by inferring them on a test sample. \\
We compared the performances, and concluded that the ensemble learners to be selected are formed by:
\begin{itemize}
	\item  a bagging of NN2 models for Dataset 1 at stage 20\%.
	\item a bagging of NB and k-NN models for Dataset 1 at stage 50\%.
	\item  a bagging of NB, k-NN, LR, NN2, and SVM-RBF for Dataset 2 at stage 20\%.
	\item  a bagging of LR models for Dataset 2 at stage 50\%.
\end{itemize}
whereas it was not possible to select a good ensemble for Dataset 1 at stage 50\% as none of the ensembles was statistically significant.\\
The results are good for Dataset 2 both in terms of Averaged Gini Index and p-values, especially if we consider the issues encountered. In particular, the issues are mainly the size of Dataset 1 and the unbalanced nature of Dataset 2. In turn, this  makes the multi-class classification problem more complex. This was evident by the fact that it was impossible to find a good classifier for Dataset 1 at stage 50\% and that the performance obtained for Dataset 1 at stage 20\% was poor.\\
Based on the aforementioned research limitations, below are some suggestions for our future work:
\begin{itemize}
	\item The best way to face the dataset size issue would be to have more data available, by collecting training and testing datasets for every time the course is offered.
	\item We also suggest to perform several additional splits for Dataset 1 at Stage 20\% to check the robustness of the model as well as the statistical significance. 
	\item It might be worth trying to optimize the topology of the neural network with a dedicated algorithm. Even though our choice was based on recent literature it is unlikely that we reached the optimum.  One could consider to try, for instance, all the possible combinations with 1, 2, and 3 layers and 1,..,20 neurons in each layer. If we considered all such combinations we would have had $20 + 20^2 + 20^3$ NNs to train. Of course it would be computationally not viable and would probably result in a massive over-fitting. However, there are several approaches proven to be effective in this kind of tasks, such as genetic optimization or pre-trained models capable to predict the optimal topology of a network for a given problem, considering parameters such as the dimension of the dataset and the intensity of the noise \cite{multi_res}.
\end{itemize}
\small
\textbf{Datasets' Permissions}
\begin{itemize}
	\item Dataset 1: The dataset is publicly available at:\\ \url{https://sites.google.com/site/learninganalyticsforall/data-sets/epm-dataset}.\\ Use of this data set in publications was acknowledged by referencing \cite{67}.
	\item Dataset 2: All permissions to use this dataset were obtained through The University of Western Ontario's Research Ethics Office. This office approved the use of this dataset for research purposes.
\end{itemize}
\textbf{Acknowledgments}  This study was funded by Ontario Graduate Scholarship (OGS) Program.\\
\textbf{Conflict of Interest}  The authors declare that they have no conflict of interest.\\
\textbf{Informed Consent}  This study does not involve any experiments on animals.

%
%

\bibliographystyle{spbasic}      
\bibliography{ref}   


\end{document}